\documentclass[longauth]{aa} % add referee option for referee mode
\usepackage[varg]{txfonts} 
\usepackage{graphicx}
\usepackage{natbib}
\bibpunct{(}{)}{;}{a}{}{,} 
\setcitestyle{notesep={ }} % removes comma between date and post-note in reference
\usepackage{multirow}
\usepackage{hyperref}
\hypersetup{
colorlinks,
citecolor=blue,
}
\usepackage{xcolor} % highlight text in color with \textcolor{color}{text}

\DeclareMathOperator*{\median}{median}
\defcitealias{HC21}{HC21} 

\begin{document}

\title{The miniJPAS survey:\\ 
Maximising the photo-$z$ accuracy from multi-survey datasets\\
with probability conflation} 
\author{A.~Hern\'an-Caballero\inst{\ref{CEFCA2} \thanks{email: ahernan@cefca.es}}
\and M.~Akhlaghi\inst{\ref{CEFCA2}}
\and C.~L\'opez-Sanjuan\inst{\ref{CEFCA2}}
\and H.~V\'azquez Rami\'o\inst{\ref{CEFCA2}}
\and J.~Laur\inst{\ref{Tartu}}
\and J.~Varela\inst{\ref{CEFCA}}
\and T.~Civera\inst{\ref{CEFCA}}
\and D.~Muniesa\inst{\ref{CEFCA}}
\and A.~Finoguenov\inst{\ref{UH}}
\and J.~A.~Fern\'andez-Ontiveros\inst{\ref{CEFCA2}}
\and H.~Dom\'inguez S\'anchez\inst{\ref{CEFCA2}}
\and J.~Chaves-Montero\inst{\ref{IFAE}}
\and A.~Fern\'andez-Soto\inst{\ref{IFCA},\ref{IFCA-UV}}
\and A.~Lumbreras-Calle\inst{\ref{CEFCA}}
\and L.A.~D\'iaz-Garc\'ia\inst{\ref{IAA}}
\and A.~del Pino\inst{\ref{CEFCA2}}
\and R.M.~Gonz\'alez Delgado\inst{\ref{IAA}}
\and C.~Hern\'andez-Monteagudo\inst{\ref{ULL},\ref{IAC}}
\and P.~Coelho\inst{\ref{SaoPaulo2}}
\and Y.~Jim\'enez-Teja\inst{\ref{IAA},\ref{ON}}
\and P.A.A.~Lopes\inst{\ref{Valongo}}
\and V.~Marra\inst{\ref{NucleoBR},\ref{INAF},\ref{IFPU}}
\and E.~Tempel\inst{\ref{Tartu}}
\and J.M.~V\'ilchez\inst{\ref{IAA}}
\and R.~Abramo\inst{\ref{IF/USP}}
\and J.~Alcaniz\inst{\ref{ON}}
\and N.~Ben\'itez\inst{\ref{IAA}}
\and S.~Bonoli\inst{\ref{DIPC},\ref{Iker}}
\and S.~Carneiro\inst{\ref{IF/UFB}}
\and J.~Cenarro\inst{\ref{CEFCA2}}
\and D.~Crist\'obal-Hornillos\inst{\ref{CEFCA}}
\and R.~Dupke\inst{\ref{ON},\ref{UMich},\ref{UAlab}}
\and A.~Ederoclite\inst{\ref{CEFCA2}}
\and A.~Mar\'in-Franch\inst{\ref{CEFCA2}}
\and C.~Mendes de Oliveira\inst{\ref{SaoPaulo}}
\and M.~Moles\inst{\ref{CEFCA},\ref{IAA}}
\and L.~Sodr\'e Jr.\inst{\ref{SaoPaulo}}
\and K.~Taylor\inst{\ref{Inst4}}
}
\institute{Centro de Estudios de F\'isica del Cosmos de Arag\'on (CEFCA), Unidad Asociada al CSIC, Plaza San Juan, 1, E-44001 Teruel, Spain\label{CEFCA2}
\and Tartu Observatory, University of Tartu, Observatooriumi 1, 61602 T\~oravere, Estonia\label{Tartu}
\and Centro de Estudios de F\'isica del Cosmos de Arag\'on (CEFCA), Plaza San Juan, 1, E-44001 Teruel, Spain\label{CEFCA}
\and Department of Physics, University of Helsinki, Gustaf H\:allstr\:omin katu 2, 00014 Helsinki, Finland\label{UH}
\and Institut de Física d'Altes Energies, The Barcelona Institute of Science and Technology, Campus UAB, E-08193 Bellaterra (Barcelona), Spain\label{IFAE}
\and Instituto de F\'{\i}sica de Cantabria (Consejo Superior de Investigaciones Cient\'{\i}ficas – Universidad de Cantabria). Avda. de los Castros, 39005, Santander, Spain\label{IFCA}
\and Unidad Asociada “Grupo de astrof\'{\i}sica extragal\'actica y cosmolog\'{\i}a” (IFCA – Universitat de Val\`encia). Parc Cient\'{\i}fic UV, 46980, Paterna, Spain\label{IFCA-UV}
\and Instituto de Astrof\'{\i}sica de Andaluc\'{\i}a (CSIC), P.O.~Box 3004, E-18080 Granada, Spain\label{IAA}
\and Departamento de Astrof\'isica, Universidad de La Laguna, E-38206, La Laguna, Tenerife, Spain\label{ULL}
\and Instituto de Astrof\'isica de Canarias, E-38200 La Laguna, Tenerife, Spain\label{IAC} 
\and Universidade de S\~ao Paulo, Instituto de Astronomia, Geof\'isica e Ci\^encias Atmosf\'ericas, Rua do Mat\~ao 1226, 05508-090 S\~ao Paulo, Brazil\label{SaoPaulo2}
\and Observat\'orio Nacional, Minist\'erio da Ci\^encia, Tecnologia, Inova\c{c}\~ao e Comunica\c{c}\~oes, Rua General Jos\'e Cristino, 77, S\~ao Crist\'ov\~ao, 20921-400, Rio de Janeiro, Brazil\label{ON}
\and Observat\'orio do Valongo, Universidade Federal do Rio de Janeiro, Ladeira do Pedro Ant\^onio 43, Rio de Janeiro, RJ, 20080-090, Brazil\label{Valongo}
\and N\'ucleo de Astrof\'isica e Cosmologia \& Departamento de F\'isica, Universidade Federal do Esp\'irito Santo, 29075-910, Vit\'oria, ES, Brazil\label{NucleoBR}
\and INAF -- Osservatorio Astronomico di Trieste, via Tiepolo 11, 34131 Trieste, Italy\label{INAF}
\and IFPU -- Institute for Fundamental Physics of the Universe, via Beirut 2, 34151, Trieste, Italy\label{IFPU}
\and Instituto de F\'isica, Universidade de S\~ao Paulo, Rua do Mat\~ao 1371, CEP 05508-090, S\~ao Paulo, Brazil\label{IF/USP}
\and Donostia International Physics Centre, Paseo Manuel de Lardizabal 4, E-20018 Donostia-San Sebastian, Spain\label{DIPC}
\and Ikerbasque, Basque Foundation for Science, E-48013 Bilbao, Spain\label{Iker}
\and Instituto de F\'isica, Universidade Federal da Bahia, 40210-340, Salvador, BA, Brazil\label{IF/UFB}
\and Department of Astronomy, University of Michigan, 311 West Hall, 1085 South University Ave., Ann Arbor, USA\label{UMich}
\and Department of Physics and Astronomy, University of Alabama, Box 870324, Tuscaloosa, AL, USA\label{UAlab}
\and Departamento de Astronomia, Instituto de Astronomia, Geof\'isica e Ci\^encias Atmosf\'ericas da USP, Cidade Universit\'aria, 05508-900, S\~ao Paulo, SP, Brazil\label{SaoPaulo}
\and Instruments4, 4121 Pembury Place, La Ca\~nada-Flintridge, CA, 91011, USA\label{Inst4}
}

\date{Accepted ........ Received ........;}

\abstract {
We present a new method for obtaining photometric redshifts (photo-$z$) for sources observed by multiple photometric surveys using a combination (conflation) of the redshift probability distributions (PDZs) obtained independently from each survey. The conflation of the PDZs has several advantages over the usual method of modelling all the photometry together, including the modularity, speed, and accuracy of the results. 
Using a sample of galaxies with narrow-band photometry in 56 bands from J-PAS and deeper $grizy$ photometry from the Hyper-SuprimeCam Subaru Strategic program (HSC-SSP), we show that PDZ conflation significantly improves photo-$z$ accuracy compared to fitting all the photometry or using a weighted average of point estimates.
The improvement over J-PAS alone is particularly strong for $i$$\gtrsim$22 sources, which have low signal-to-noise ratios in the J-PAS bands. For the entire $i$$<$22.5 sample, we obtain a 64\% (45\%) increase in the number of sources with redshift errors $\vert\Delta z\vert$$<$0.003, a factor of 3.3 (1.9) decrease in the normalised median absolute deviation of the errors ($\sigma_{\rm{NMAD}}$), and a factor of 3.2 (1.3) decrease in the outlier rate ($\eta$) compared to J-PAS (HSC-SSP) alone.
The photo-$z$ accuracy gains from combining the PDZs of J-PAS with a deeper broad-band survey such as HSC-SSP are equivalent to increasing the depth of J-PAS observations by $\sim$1.2--1.5 magnitudes. 
These results demonstrate the potential of PDZ conflation and highlight the importance of including the full PDZs in photo-$z$ catalogues.
}

\keywords{surveys - techniques:photometric - methods: data analysis - galaxies: distances and redshifts}

\titlerunning{Maximising photo-$z$ accuracy with conflation}
\authorrunning{A. Hern\'an-Caballero et al.}

\maketitle

\section{Introduction}

In this decade, a number of current and upcoming large-area imaging surveys such as the Rubin Observatory Legacy Survey of Space and Time \citep[LSST;][]{LSST-SC09}, the Dark Energy Survey \citep[DES;][]{DES16}, the Hyper Suprime-Cam Subaru Strategic Program \citep[HSC-SSP;][]{Aihara18}, and the Kilo-Degree Survey \citep[KiDS;][]{Hildebrandt21} will cover thousands of square degrees in the sky to depths that previously were only possible for small surveys of a few square degrees at most. 
These surveys use sets of broad-band filters that are carefully designed to maximise their performance for photometric redshifts (photo-$z$), allowing for new applications in galaxy evolution and cosmology \citep[e.g. see][and references therein]{Newman22}. 

The Sloan Digital Sky Survey \citep[SDSS;][]{York00} showed that five broad bands ($ugriz$) suffice to obtain relative errors in photo-$z$, $\Delta z$ = ($z_{\rm{phot}}$-$z_{\rm{spec}}$)/(1+$z_{\rm{spec}}$), as low as $\sigma$($\Delta z$) $\sim$ 2\% for sources meeting some magnitude and colour cuts \citep{Beck16}. A similar accuracy level has been reported for much fainter galaxies observed by HSC-SSP ($grizy$ bands) over a wide range of redshifts \citep{HSC-collaboration23}.
However, some applications, such as baryonic acoustic oscillation (BAO) measurements, require higher photo-$z$ accuracy, with uncertainties as low as $\sigma$($\Delta z$) $\sim$ 0.3--0.5\% \citep{Blake05,Angulo08,Benitez09,Chaves-Montero18}. This level of accuracy is only achievable with narrower pass bands that provide a higher effective spectral resolution. 

The COMBO-17 \citep{Wolf03}, Subaru COSMOS 20 \citep{Taniguchi07,Taniguchi15}, and ALHAMBRA \citep{Moles08} surveys achieved $\sigma$($\Delta z$) $\lesssim$ 1\% with a combination of broad-band filters and medium-band filters with full width at half maximum (FWHM) of $\sim$300 \AA. 
More recently, \citet{Barro19} reported that, for galaxies in the GOODS-North field with multi-wavelength (UV to infrared) broad-band observations, the photo-$z$ accuracy improves eightfold (from $\sigma$($\Delta z$) $\sim$ 2.3\% to $\sigma$($\Delta z$) $\sim$ 0.3\%) with the addition of narrow-band photometry (FWHM $\sim$ 170 \AA) in 25 optical bands from the SHARDS survey \citep{Perez-Gonzalez13}. 
Similarly, \citet{Alarcon21} obtained $\sigma$($\Delta z$) $\sim$ 0.3\% in the COSMOS field with the addition of photometry in 40 narrow bands (FWHM $\sim$ 130 \AA) from PAUS \citep{Eriksen19}.

An important step in determining accurate photo-$z$ over a large area will be taken by the Javalambre-Physics of the Accelerating Universe Astrophysical Survey \citep[J-PAS;][]{Benitez09,Benitez14}, which will cover one third of the northern hemisphere with a unique set of 56 optical filters (plus $i$ band for detection) and provides, for each 0.48\arcsec{} $\times$ 0.48\arcsec{} pixel in the sky, an R$\sim$60 photo-spectrum (J-spectrum) covering the 3800--9100 \AA{} range.
Two small surveys were carried out to demonstrate the scientific potential of J-PAS: miniJPAS \citep{Bonoli21} and J-NEP \citep{Hernan-Caballero23}, showing that $\sigma$($\Delta z$)=0.3\% is attainable with the 56 bands of J-PAS plus $u$, $g$, $r$, and $i$ \citep{HC21,Hernan-Caballero23,Laur22}.

Nevertheless, covering the optical range in many narrow-band filters is very expensive in terms of telescope time. Thus, the limiting magnitude of J-PAS \citep[$m_{5\sigma}$$\sim$22.5;][]{Bonoli21} is shallower than other large-area ($>$1000 deg$^2$) broad-band surveys such as KiDS ($\sim$1350 deg$^2$, $m_{5\sigma}$$\sim$25) and HSC-SSP Wide ($\sim$1400 deg$^2$, $m_{5\sigma}$$\sim$26). 
The results from miniJPAS and J-NEP, as well as previous results from SHARDS and PAUS, indicate that J-PAS photo-$z$ can be improved with the addition of deeper photometry from  upcoming broad-band surveys that will cover the J-PAS footprint, such as Euclid \citep[$I$$\sim$26.2, $Y$, $J$, $H$ $\sim$ 24.5][]{EuclidCollaboration22} and the China Space Station Telescope Optical Survey \citep[CSS-OS; $NUV$, $u$, $g$, $r$, $i$, $z$, $y$,  $\sim$25][]{Gong19}.

Computing accurate photo-$z$ by mixing the photometry of a shallow multi-narrow-band survey and a deep broad-band survey is technically challenging. The two main roadblocks are systematics in the photometry and the choice of the photo-$z$ method. 

The broad- and narrow-band observations are obtained with different telescopes or instruments and processed by different pipelines. This implies that the pixel scale and the point spread function (PSF) of the images might differ, as well as the methods for flux calibration and source extraction (including aperture definition), resulting in complex systematic offsets between the broad- and narrow-band fluxes. This problem is exacerbated if the broad-band survey is much deeper because the small nominal photometric uncertainties do not account for the cross-calibration uncertainty.
Even if one is to reprocess all the data with the same pipeline, obtaining consistent photometry in all the bands requires convolving all the images to the worst PSF or deblending the emission of sources in wider-PSF images using a sharper one as reference \citep[see][]{Barro19}.

Another significant limitation comes from the photo-$z$ techniques employed. For observations including only a few broad bands, the best results are obtained with an empirical method \citep[e.g.][]{Beck16,HSC-collaboration23} that uses a large sample of galaxies with spectroscopic redshifts to calibrate the dependence of the observed colours with redshift. However, the high dimensionality of the colour space that results from having a large number of narrow bands makes the empirical method unfeasible with the small training samples currently available. 
As a consequence, photo-$z$ measurements from multi-narrow-band photometry rely almost exclusively on SED-fitting with spectral templates \citep[e.g.][]{Wolf03,Moles08,Perez-Gonzalez13,Barro19,Eriksen19,Alarcon21,Laur22}.
Doing an SED fit to the combined deep and shallow photometry is tricky because of the cross-calibration issue mentioned above, but also because the optimal number of templates depends on the number of bands with detections (Hern\'an-Caballero et al. in prep.). Bright sources with strong detections in the narrow bands benefit from a larger number of templates since that makes it easier to find a good match. However, for faint sources detected only in the broad bands, such diversity of spectral shapes only adds degeneracy to the colour-redshift space, with results improving when just a few templates (representing broad spectral classes) are used \citep[e.g. see][]{Benitez00}.

An alternative to using the combined photometry from two distinct datasets is to obtain redshift probability distribution functions (PDZs) separately from the broad- and narrow-band surveys and then combine the PDZs.
To our knowledge, this approach has not been tested so far. Conceptually similar strategies were proposed by \citet{Kovac10} and \citet{Barro19} to improve the confidence of spectroscopic redshifts from single-line detections using photo-$z$ as priors.

Combining PDZs instead of photometry has some clear advantages: i) it eliminates the need to obtain consistent photometry through all the surveys; ii) it allows the use of the most suitable photo-$z$ method for each dataset; iii) it makes it possible to re-use the photo-$z$ already published by other surveys, capitalising on the expertise of the teams that produced them and making it easier and faster to add new datasets to the photo-$z$ calculation. 

In this paper, we use a sample of galaxies in the AEGIS field to show that a combination of PDZs from narrow-band (miniJPAS) and broad-band (HSC-SSP) observations results in significantly improved photo-$z$ accuracy compared to the individual surveys and also compared to SED-fitting all the photometry together.
The structure of the paper is as follows. Section \ref{sec:methods} describes the probability conflation method for combining PDZs. Section \ref{sec:sample} presents the sample of galaxies with J-PAS and HSC-SSP observations. Section \ref{sec:single-survey} discusses the photo-$z$ obtained with different codes using the two datasets separately. Section \ref{sec:combination} describes and compares three methods for obtaining photo-$z$ from the two datasets: mixing the photometry, weighted averaging of point estimates, and conflating the PDZs. Finally, Sect. \ref{sec:discussion} compares the performance of J-PAS, HSC-SSP, and their combined photo-$z$ as a function of the source magnitude and discusses the implications for J-PAS.
All magnitudes are expressed in the AB system. 

\section{Methods}\label{sec:methods}

\begin{figure}
\begin{center}
\includegraphics[width=8.4cm]{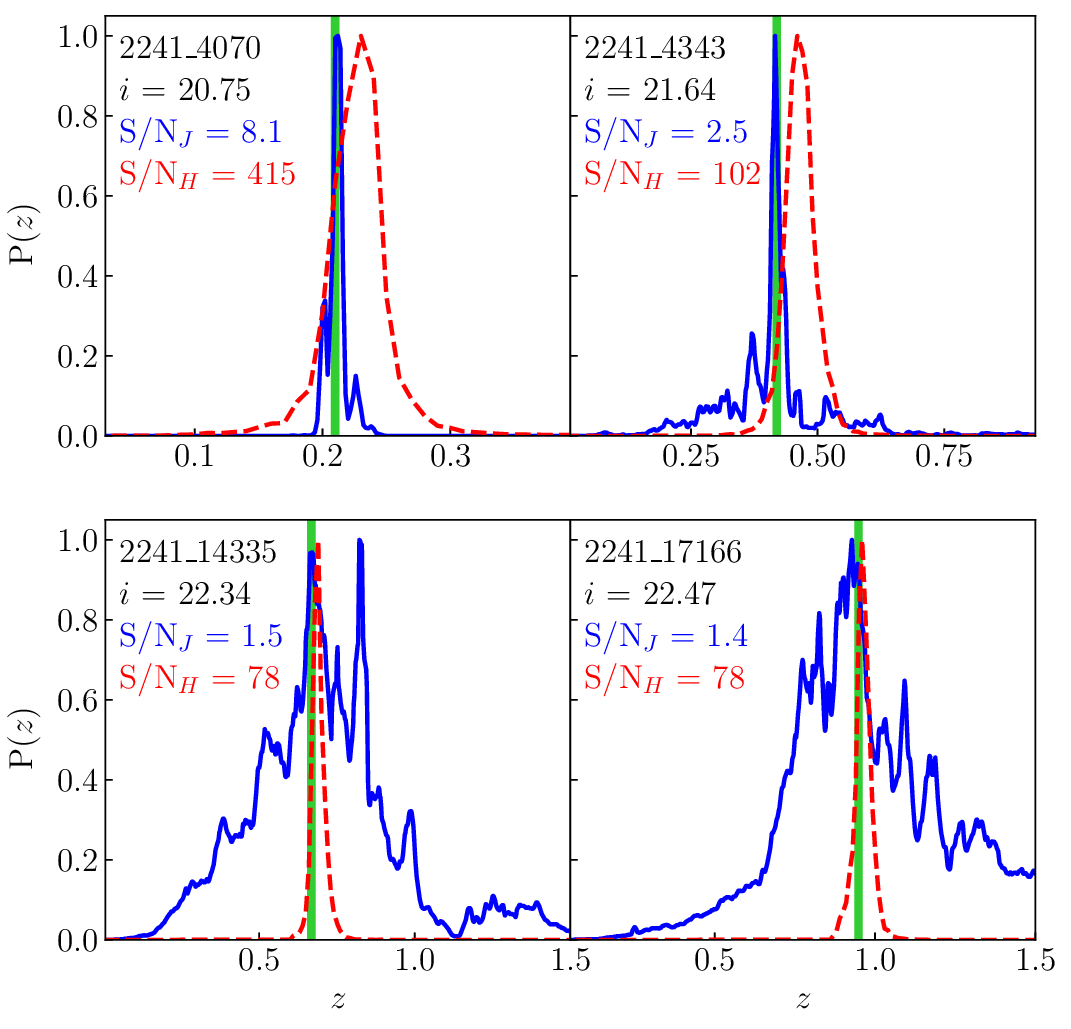}
\end{center}
\caption[]{Comparison of redshift probability distributions obtained from narrow-band J-PAS photometry (blue solid line) and deep broad-band photometry from HSC-SSP (red dashed line) in four galaxies from the AEGIS field sample. The spectroscopic redshift is marked with a green vertical line. Annotations in the top left corner of each panel indicate (from top to bottom) the source ID in the miniJPAS public data release catalogue, its $i$-band magnitude, and the median S/N per band in the J-PAS photometry and the HSC-SSP photometry.\label{fig:pdf-examples}}
\end{figure}

Let \textbf{X} = \{$X_i$\} and \textbf{Y} = \{$Y_j$\} be two independent sets of observations of the same galaxy, and $P_X$($z$) and $P_Y$($z$) the probability density distributions for the redshift derived from each of them. 
If we assume $P_X$($z$) and $P_Y$($z$) to be independent, the combined probability distribution can be expressed as the conflation of the two probability distributions \citep{Hill11}:
\begin{equation}\label{eq:conflation}
P_C(z) = \frac{P_{X}(z) \cdot P_{Y}(z)}{\int_{-\infty}^{\infty} P_{X}(z) P_{Y}(z) dz} ,
\end{equation}
\noindent where the integral in the denominator ensures that $P_C$($z$) is unitary. Conflation has multiple advantages over other methods of combining probability distributions, including the following: minimisation of the loss of Shannon information, being a best linear unbiased estimate, and yielding a maximum-likelihood estimator \citep{Hill11}.

The main disadvantage of the conflation method is that it is sensitive to over-confidence in the individual PDZs. Over-confidence (defined as confidence intervals containing the true redshift less often than predicted from the PDZs, implying the PDZs are too narrow) is a common issue in many photo-$z$ codes \citep[see][for a review]{Dahlen13}. Over-confidence in one or both PDZs may cause only their wings to overlap or, in extreme cases, not to overlap at all, resulting in an undefined $P_C$($z$). Therefore, successful conflation requires the PDZs to be realistic or under-confident. 

In a Bayesian framework, the probability $P_X$($z$) can be written as
\begin{equation}
P_X(z) \propto \int_{\boldsymbol{\theta}} \mathcal{L}(\mathbf{X} \vert z, \boldsymbol{\sigma}, \boldsymbol{\theta}) P(z, \boldsymbol{\theta}) d\boldsymbol{\theta} ,
\end{equation}
\noindent where $\mathcal{L}$(\textbf{X} $\vert$ $z$, $\boldsymbol{\sigma}$, $\boldsymbol{\theta}$) is the likelihood of the observations \textbf{X} given the uncertainties $\boldsymbol{\sigma}$ = \{$\sigma_i$\} and the parameters $\boldsymbol{\theta}$ = \{$\theta_k$\} that define the model\footnote{These parameters may represent the template set in the case of the SED-fitting method or a parametrisation of the colour-redshift space in the case of the empirical method.}. $P$($z$, $\boldsymbol{\theta}$) is the prior on $z$ and $\boldsymbol{\theta}$, which represent our knowledge (obtained elsewhere) about the distribution of redshift and spectral properties for galaxies of a given magnitude ($N$($z$,$\boldsymbol{\theta}$|$m$), hereafter N($z$)).

Since the prior does not depend on the observations \textbf{X} or \textbf{Y}, $P_X$($z$) and $P_Y$($z$) are not entirely independent. This results in $P_C$($z$) being overconfident because the prior information is considered twice.
The impact of the prior is stronger if $\mathcal{L}$ is unable to tightly constrain $z$ and $\boldsymbol{\theta}$ (e.g. noisy or scarce data). To prevent this prior-related over-confidence, we used the N($z$) prior for the computation of $P_X$($z$) but imposed a flat one ($P$($z$,$\boldsymbol{\theta}$) = 1) for $P_Y$($z$), which does not favour any particular redshift value over others. This is conceptually equivalent to using $P_X$($z$) as the prior in the calculation of $P_Y$($z$) from $\mathcal{L}$(\textbf{Y} $\vert$ $z$, $\boldsymbol{\sigma}$, $\boldsymbol{\theta}$):
\begin{equation}
P_C(z) = P^*_Y(z) \propto P_X(z) \int_{\boldsymbol{\theta}} \mathcal{L}(\mathbf{Y} \vert z, \boldsymbol{\sigma}, \boldsymbol{\theta}) d\boldsymbol{\theta} .
\end{equation}
The same method can be extended to combine the probability distributions from $n$ datasets:
\begin{equation}
P_C(z) \propto P_1(z) \cdot P_2^-(z) \cdots P_n^-(z) ,
\end{equation}
\noindent where the probabilitites $P_i^-$($z$) for $i$ = 2, \dots, $n$ are computed with a flat prior.

Conflation can be particularly useful to combine PDZs from shallow narrow-band observations with deeper ones in broad bands due to how the spectral resolution ($R$) and the signal-to-noise ratio (S/N) of the observations impact the shape of the PDZ. The width $\delta z$ of the peaks in $\mathcal{L}$($z$) is proportional to the width $\delta \lambda$ of the spectral features that can be resolved by the photometry:
\begin{equation}
\frac{\delta z}{1+z} \propto \frac{\delta \lambda}{\lambda} = \frac{1}{R} .
\end{equation}
Therefore, everything else being equal, a factor of two reduction in the FWHM of the filters roughly translates into a factor of two decrease in photo-$z$ errors.
On the other hand, the S/N of the observations impacts the number and contrast of the peaks in $\mathcal{L}$($z$). At high S/N, $\mathcal{L}$($z$) is usually unimodal since the observed colours are only compatible with a reduced range of intrinsic colours and redshifts\footnote{Unless the number of bands or the spectral coverage is insufficient to break intrinsic degeneracies in the colour-redshift space; see \citet{Benitez00} for a discussion.}. As the S/N per band decreases, the larger photometric uncertainties can make the observations compatible with combinations of intrinsic colours and redshift that are very different from the actual ones, resulting in spurious peaks in $\mathcal{L}$($z$) located far from the true redshift of the galaxy. At S/N $\lesssim$ 1, the number of peaks in $\mathcal{L}$($z$) increases dramatically, while the contrast between peaks and valleys decreases, resulting in a roughly flat distribution with minor fluctuations. At this point, $\mathcal{L}$($z$) contains very little information, and the PDZ becomes dominated by the prior. 

Figure \ref{fig:pdf-examples} shows representative examples of PDZs obtained from miniJPAS observations (using only the narrow bands) and from the much deeper broad-band observations of HSC-SSP, with the S/N per band $\sim$25--30 times higher (see Sect. \ref{sec:sample}). 
For relatively bright sources ($i$$\lesssim$21) with median S/N $\gtrsim$3 in the narrow bands, the J-PAS PDZ usually has a single peak concentrating most of the probability density. This peak is narrower than the PDZ from HSC-SSP despite the much higher S/N because its spectral resolution, not sensitivity, limits the latter. 
For fainter sources ($i$$\gtrsim$22), the PDZ from HSC-SSP is still unimodal and has the same width, while the PDZ of J-PAS is now much broader since it is composed of multiple overlapping peaks whose intensity is modulated by the prior. This implies that point estimates, such as the mode of the PDZ, can be highly inaccurate (an outlier) if photometric errors or the prior favour the wrong peak. In fact, the broad and multi-modal PDZs of faint J-PAS sources result in a high rate of outliers (see Sects. \ref{sec:photoz-miniJPAS} and \ref{sec:discussion}).

Throughout this paper, we use several quantities related to photometric redshifts and their accuracy. While some are relatively standard, others are not. Here, we define all of them for convenience:

\begin{itemize}
\item{$z_{phot}$ : A generic term for any point estimate of the photometric redshift. It can be obtained from the PDZ in several ways, such as taking the mode, the mean, the median, or a randomly sampled value. In this paper, $z_{phot}$ represents the mode of the PDZ unless otherwise indicated.}\\

\item{$\Delta z$ : The relative error in $z_{phot}$, defined as
\begin{equation}
\Delta z = (z_{phot}-z_{spec})/(1+z_{spec}) .
\end{equation}}

\item{$odds$ : An indicator of the confidence in $z_{phot}$. It represents the probability that $\vert \Delta z \vert$ is smaller than a given threshold $\delta$. In this work, we used $\delta$ = 0.03. The $odds$ value is computed by integration of the PDZ as
\begin{equation}
odds = \int_{z_{phot}-\delta(1+z_{phot})}^{z_{phot}+\delta(1+z_{phot})} P(z) dz .
\end{equation}}

\item{$\eta$ : The fraction of outliers in a sample (also called the outlier rate) where an `outlier' is defined as a source with a relative error in $z_{phot}$ larger than a given threshold ($\vert\Delta z\vert$$>$$D$). In this work, we mainly used $D$=0.03 (relative error $>$3\%). If a different threshold was used, we indicate it with a sub-index (e.g.: $\eta_{15}$ for $D$=0.15 or relative error $>$15\%, the threshold most often used in papers discussing photometric redshifts from broad-band data).}\\

\item{$\sigma$($\Delta z$) : The standard deviation of $\Delta z$ in a sample.}\\

\item{$\sigma_{\rm{NMAD}}$ : A robust statistic equivalent to $\sigma$($\Delta z$), but less sensitive to outliers. It takes the same value of $\sigma$($\Delta z$) if the distribution of $\Delta z$ is Gaussian. It is defined as 
\begin{equation}
\sigma_{\rm{NMAD}} = 1.48 \times \median_i \vert \Delta z_i - \median_i(\Delta z_i)\vert , 
\end{equation}
\noindent where the index $i$ runs over the sources in the sample.
}

\item{$f_{K}$ : The fraction of sources in a sample with relative errors in $z_{phot}$ smaller than a threshold $K$. We mainly used $f_{03}$ to represent the fraction of sources with $\vert\Delta z\vert$$<$0.003 (0.3\%), a threshold relevant to BAO studies. We also provide $f_{1}$ (fraction with relative error $<$1\%), which is more relevant to the measurement of galaxy properties.}
\end{itemize}

\section{Sample selection}\label{sec:sample}

\begin{figure} 
\begin{center}
\includegraphics[width=8.4cm]{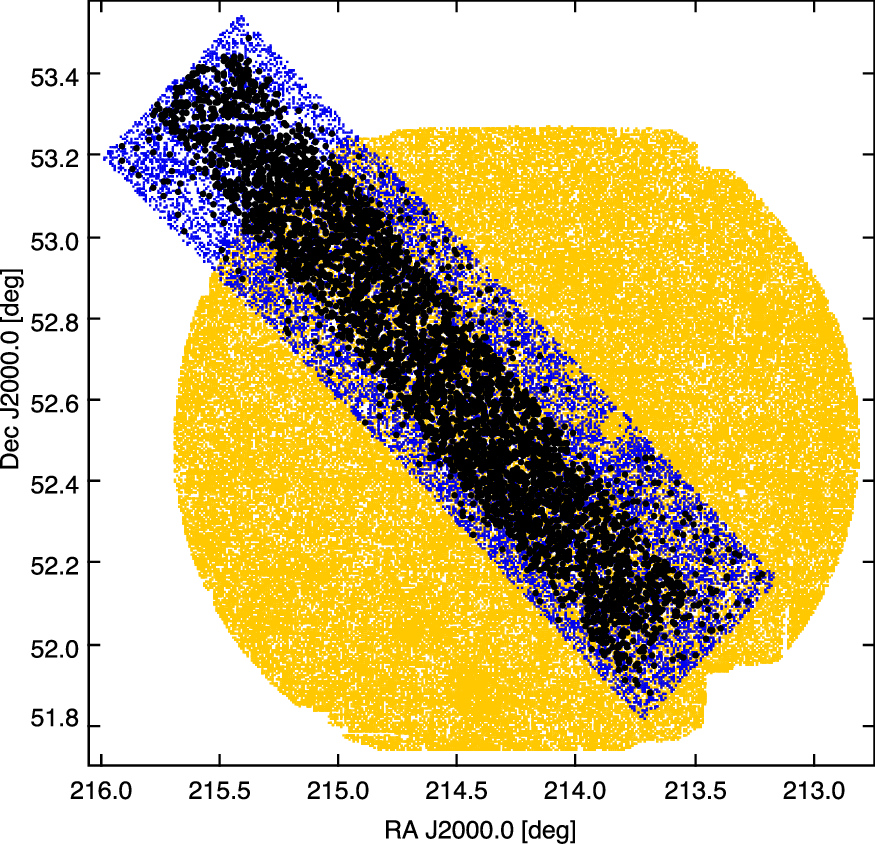}
\end{center}
\caption[]{Footprint of multi-band photometry observations in the AEGIS field. Orange dots represent $i$$<$23.5 sources in the SUBARU HSC-SSP PDR3 catalogue with photometry in the $grizy$ bands. Blue dots represent $i$$<$22.5 sources in the miniJPAS PDR201912 catalogue. Black dots mark the miniJPAS sources with spectroscopic redshift.\label{fig:footprint}}
\end{figure}

\begin{figure} 
\begin{center}
\includegraphics[width=8.4cm]{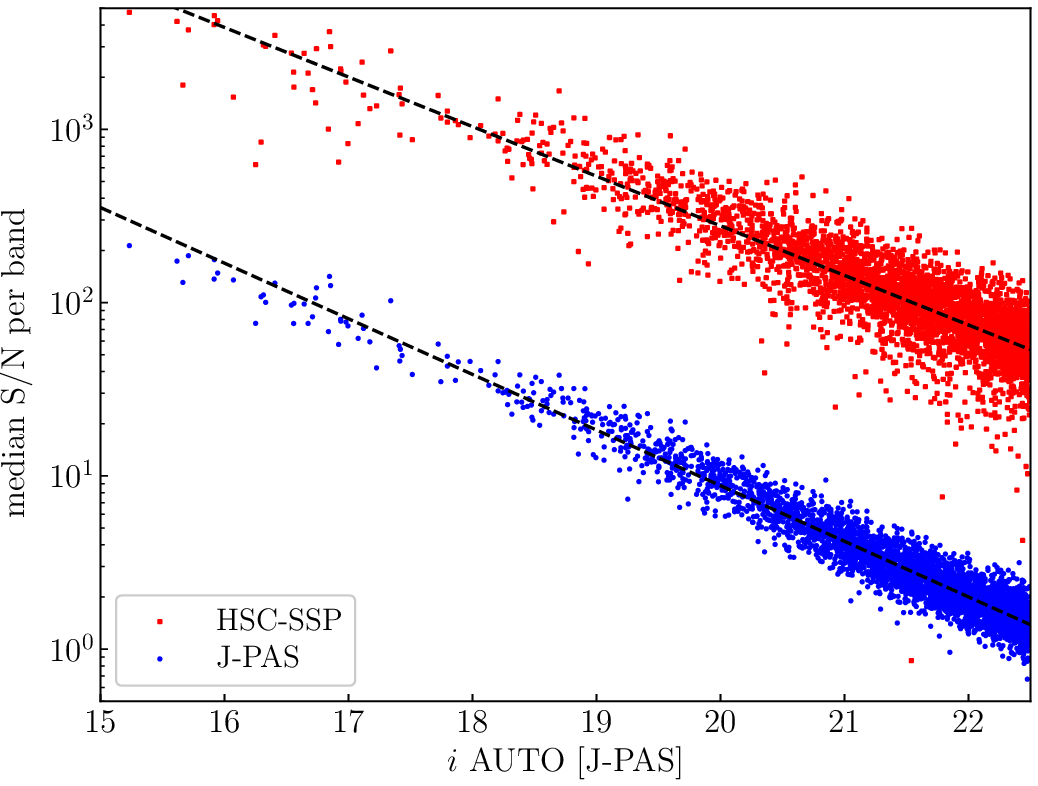}
\end{center}
\caption[]{Comparison of median S/N per band in the narrow-band photometry of miniJPAS (blue dots) and the broad-band photometry of HSC-SSP (red dots) as a function of the $i$-band magnitude of the galaxies. Dashed lines show the best-fitting log-linear regression model.\label{fig:snr-imag}}
\end{figure}

\begin{figure*}
\begin{center}
\includegraphics[width=18.0cm]{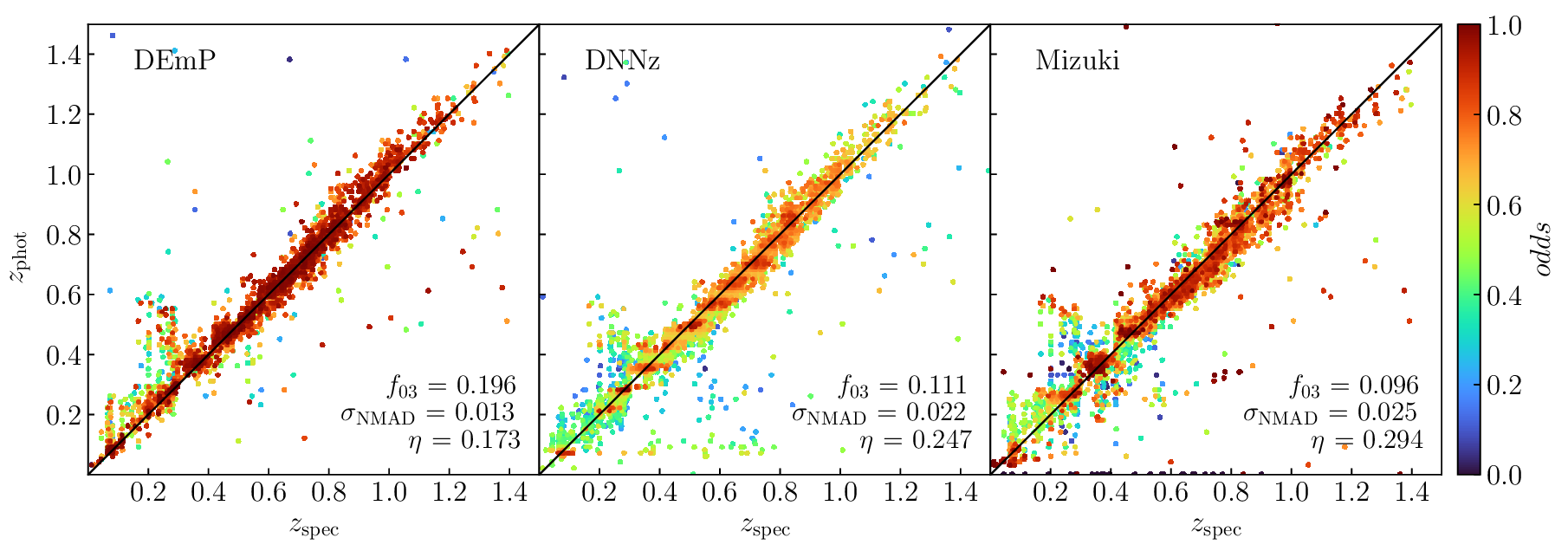}
\end{center}
\caption[]{Comparison of photometric versus spectroscopic redshift for the individual galaxies in our sample. Each panel shows the results from a different photo-$z$ code using the photometry from HSC-SSP in the $grizy$ bands. Symbols are colour-coded for the $odds$ parameter.\label{fig:zphot-zspec-SUBARU}}
\end{figure*}

Our parent sample is comprised of all 21235 sources with magnitude $i$$<$22.5 in the AUTO aperture from the public data release of miniJPAS\footnote{\url{http://archive.cefca.es/catalogues/minijpas-pdr201912}}. We cross-matched this sample with a spectroscopic redshift catalogue that combines redshifts from the DEEP DR4 \citep{Newman13} and SDSS DR16 \citep{Ahumada20} surveys using a 1\arcsec search radius. 
The spectroscopic coverage from DEEP spans a narrow strip that includes $\sim$50\% of the miniJPAS footprint (Fig. \ref{fig:footprint}). Outside this strip, only a few hundred SDSS spectra are available. In total, there are 7123 sources in miniJPAS with spectroscopic redshifts and $i$$<$22.5. 

From this sample, we removed the sources for which the spectroscopic redshift is unreliable (zWarning > 0 in SDSS; ZQUALITY < 3 in DEEP), or with spectral classification other than galaxy (stars and quasars), or with unreliable miniJPAS photometry ({\sc SExtractor} flags > 0 or mask\_flags > 0; see \citealt{Bonoli21} for details). Finally, we removed four galaxies with $z_{spec}$$>$1.5, which are beyond the search range that we used to compute photo-$z$ for J-PAS galaxies (0$<$$z$$<$1.5). After these cuts, we were left with 4448 sources.

The miniJPAS footprint is partially covered by deep imaging ($\sim$26 mag at 5$\sigma$) in the $grizy$ bands from the Hyper Suprime-Cam Subaru Strategic Program \citep[HSC-SSP;][]{Aihara18}. The roughly circular footprint of HSC-SSP is larger than miniJPAS but overlaps only $\sim$80\% of its area (Fig. \ref{fig:footprint}).
Out of the 4448 miniJPAS sources selected above, 3712 are inside the HSC-SSP footprint.
To cross-match our sample with the HSC-SSP catalog, we retrieved all the sources
with \textit{cmodel} magnitude $i$$<$23.5 from the HSC-SSP database.\footnote{\url{https://hsc-release.mtk.nao.ac.jp/datasearch/}} \footnote{The fainter magnitude cut in HSC-SSP is to ensure that no sources are lost due to uncertainties in the $i$-band flux.} Inside the overlap region, all but two miniJPAS sources with spectroscopy have an HSC-SSP counterpart within 1\arcsec. 
Our final sample includes the 3710 miniJPAS sources with $i$$<$22.5, no photometry flags, reliable $z_{spec}$$<$1.5, and HSP-SSP photometry.

Figure \ref{fig:snr-imag} compares the depth of the J-PAS and HSC-SSP observations using the median S/N of all the bands for each of the sources.
The median S/N correlates with the $i$-band magnitude of the sources, with some dispersion caused by differences in their spectral type and redshift. On average, the S/N of a galaxy is $\sim$25-30 times higher in the HSC-SSP bands compared to the J-PAS bands. 

We note that the 5-$\sigma$ depth in a 3\arcsec aperture of the miniJPAS images ranges between $\sim$21.5 and $\sim$24 magnitudes, depending on the band and pointing, with an average of $\sim$22.5 \citep[see Fig. 4 in][]{Bonoli21}. However, the median S/N per band of $i$$\sim$22.5 galaxies is just $\sim$1.5 due to the faintest galaxies having mostly red spectral energy distributions.
  
\section{Single-survey photometric redshifts}\label{sec:single-survey}

In this section, we describe and compare the photo-$z$ for our sample of $i$$<$22.5 galaxies obtained independently with three different codes using the HSC-SSP photometry, and with one code (for two different configurations) using the J-PAS bands. 

\subsection{Photo-$z$ from Subaru HSC-SSP}\label{sec:photoz-subaru}

\begin{figure} 
\begin{center}
\includegraphics[width=8.4cm]{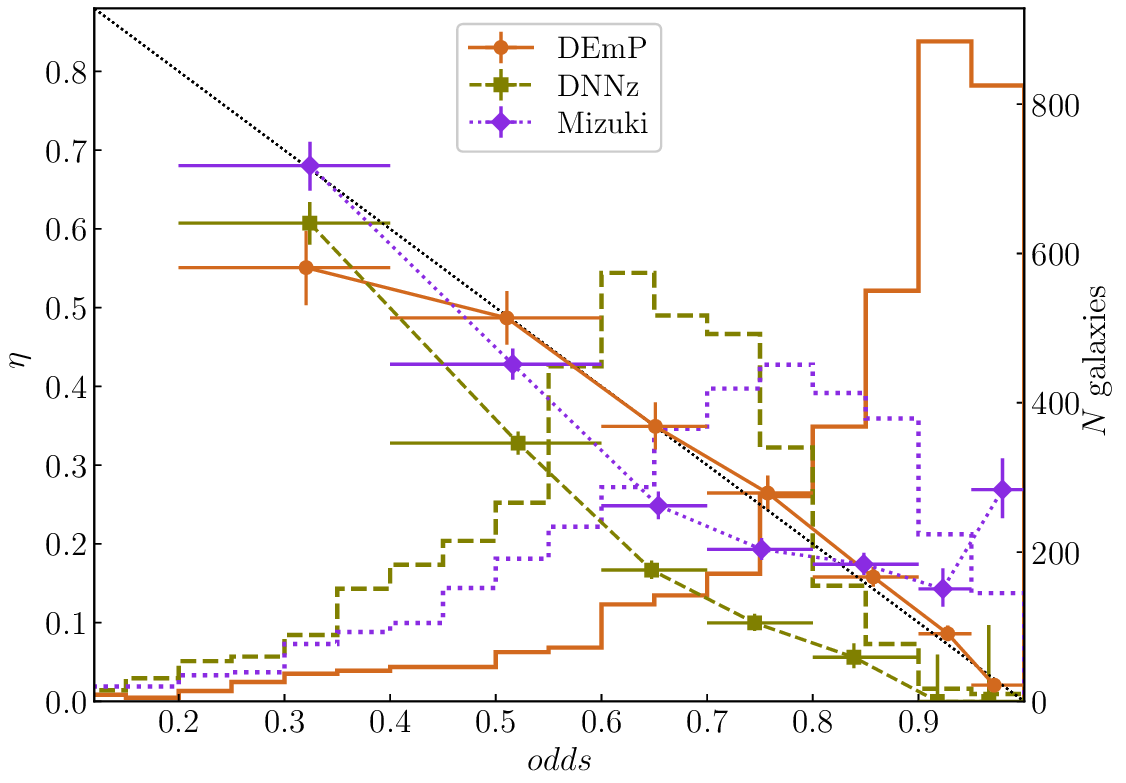}
\end{center}
\caption[]{Outlier rate as a function of the $odds$ parameter (symbols with error bars) and distribution of $odds$ values (open histograms, scale on the right axis) for photo-$z$ obtained with each of the photo-$z$ codes used by HSC-SSP. Horizontal error bars represent the intervals in which $\eta$ is calculated. Vertical error bars indicate 68\% confidence intervals for $\eta$ estimated with the \citet{Wilson27} formula for binomial distributions. The dotted diagonal line represents the expected relation between $\eta$ and $odds$ for realistic PDZs.\label{fig:outrate-odds-SUBARU}}
\end{figure}

The Public Data Release 3 \citep[PDR3;][]{Aihara22} of HSC-SSP contains photometric redshifts obtained with three different codes: i) {\sc DEmP} \citep{Hsieh14,Tanaka18}, an empirical quadratic polynomial photo-$z$ code; ii) {\sc mizuki} \citep{Tanaka15,Tanaka18}, a template-fitting code; iii) {\sc DNNz} \citep{HSC-collaboration23}, a deep learning code with a multi-layer perceptron architecture. 
We retrieved photo-$z$ from the three codes for our sample using the HSC-SSP database. Following the advice in \citet{Tanaka18},
for the codes {\sc DNNz} and {\sc mizuki} we chose the value $z_{best}$ that minimises the probability of the inferred redshift being an outlier as the default point estimate
($z_{phot}$) \citep[see][for details]{Tanaka18}. For {\sc DEmP}, we instead used the mode of the PDZ, which performs slightly better than $z_{best}$ for this particular code. 

In Fig. \ref{fig:zphot-zspec-SUBARU}, we compare the accuracy of $z_{phot}$ obtained by the three codes for the galaxies in our sample. 
The statistics $f_{03}$, $\sigma_{\rm{NMAD}}$, and $\eta$ are shown at the bottom right corner of each panel. {\sc DEmP} clearly outperforms the other two codes in these three scores (we note that, for $f_{03}$, higher is better). {\sc DEmP} is also the least affected by systematic issues at $z$$<$0.3 \citep[see][for a discussion]{HSC-collaboration23}.

While evaluating the accuracy of PDZs for individual sources is not possible, we can perform statistical tests to determine if, on average, $z_{spec}$ falls within a given confidence interval of the PDZ with the expected frequency. 
Figure \ref{fig:outrate-odds-SUBARU} shows the distribution of $odds$ and the relation between $odds$ and $\eta$ for the three codes. If the PDZs are realistic, the expected value of $\eta$ is related to the mean $odds$ by
\begin{equation}
\eta = 1 - \langle odds \rangle .
\end{equation}

This theoretical relation is represented by the dotted diagonal line in Fig. \ref{fig:outrate-odds-SUBARU}. The outlier rate for {\sc DNNz} is systematically below the expected relation (i.e. the $odds$ are under-confident). In contrast, {\sc mizuki} produces $odds$ that are  under-confident in the intermediate range 0.5 $\lesssim$ $odds$ $\lesssim$ 0.7) but over-confident in the high end ($odds$ $\gtrsim$ 0.9). Finally, the $odds$ from {\sc DEmP} are realistic in the entire range except for the few sources with $odds$$\lesssim$0.4. Additionally, {\sc DEmP} is the only code that obtains high $odds$ for a large fraction of the sample. Therefore, in the following analysis, we only focus on {\sc DEmP} results (see Appendix \ref{sec:appendix} for the results of conflation of PDZs from each of the three codes with the J-PAS likelihood).

\begin{figure} 
\begin{center}
\includegraphics[width=8.4cm]{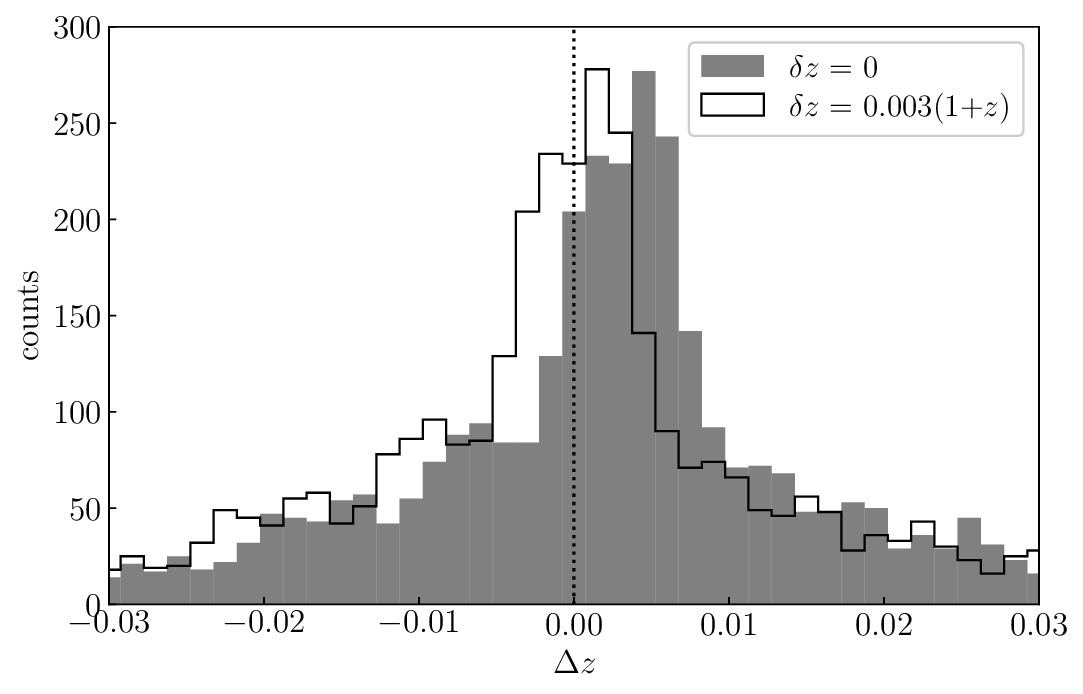}
\includegraphics[width=8.4cm]{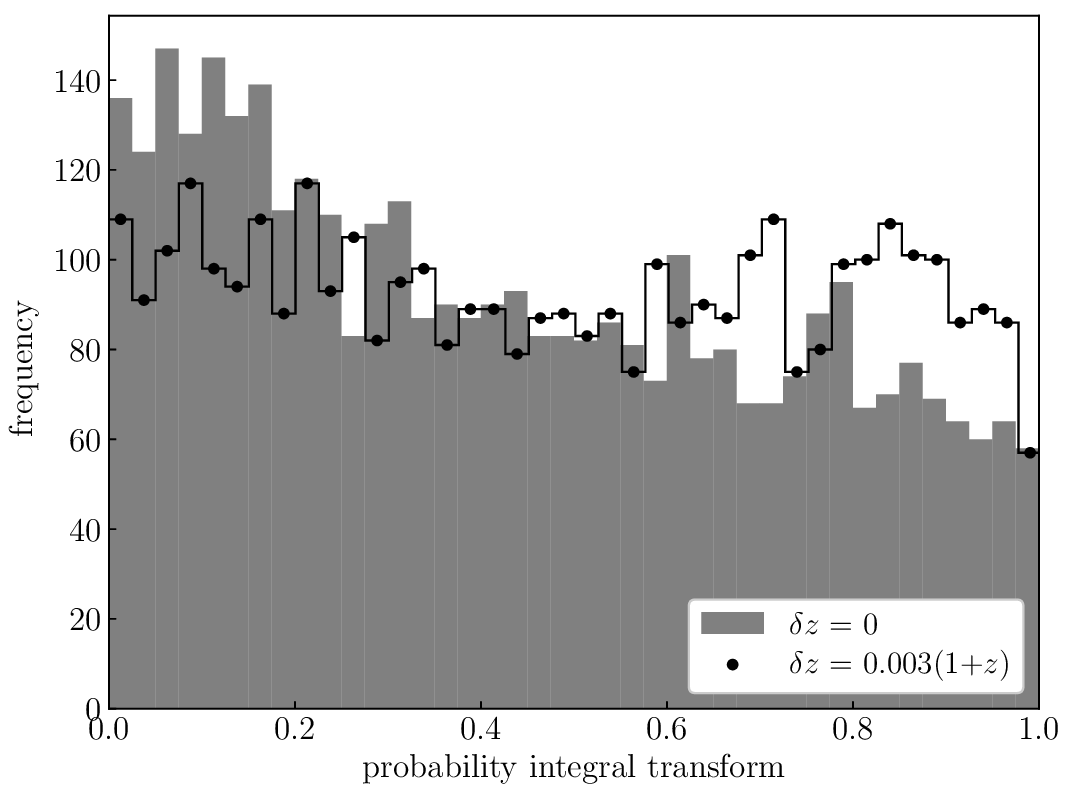}
\end{center}
\caption[]{Close-up of [-0.03,0.03] interval of the distribution of $\Delta z$ (top) and distribution of PIT values (bottom) for photo-$z$ measurements with {\sc DEmP} before (solid histograms) and after (open histograms) correcting the PDZs for the systematic offset $\delta z$=0.003(1+$z$).\label{fig:dz_offset_SUBARU}}
\end{figure}

The distribution of $\Delta z$ for {\sc DEmP} (top panel in Fig. \ref{fig:dz_offset_SUBARU}) shows that $z_{phot}$ is slightly biased towards $z_{phot}$$>$$z_{spec}$. Assuming that the bias scales with (1+$z$), we obtain $\delta z$ $\sim$ 0.003(1+$z$). After correcting for this bias, the $f_{03}$ score increases from 0.196 to 0.267, while the impact on $\sigma_{\rm{NMAD}}$ and $\eta$ is negligible.

The probability integral transform (PIT) is another test for the PDZs. The PIT values of individual sources are computed as the cumulative distribution function of the PDZ evaluated at $z_{spec}$ \citep[e.g.][]{Schmidt20}. A well-calibrated PDZ should show a flat distribution of PIT values. The PIT distribution for {\sc DEmP} PDZs (bottom panel in Fig. \ref{fig:dz_offset_SUBARU}) shows a slope in the distribution of frequencies that is also consistent with biased PDZs \citep[see][]{Polsterer16}. Shifting the PDZs by $\delta z$ $\sim$ 0.003(1+$z$) results in a roughly flat distribution. We note that if a PDZ is inconsistent with the spectroscopic redshift ($P$($z_{\rm{spec}}$) = 0), it will result in PIT=0 or PIT=1. There are no such cases in the {\sc DEmP} PDZs from our sample.

\subsection{Photo-$z$ from miniJPAS}\label{sec:photoz-miniJPAS}

\begin{figure}
\begin{center}
\includegraphics[width=8.4cm]{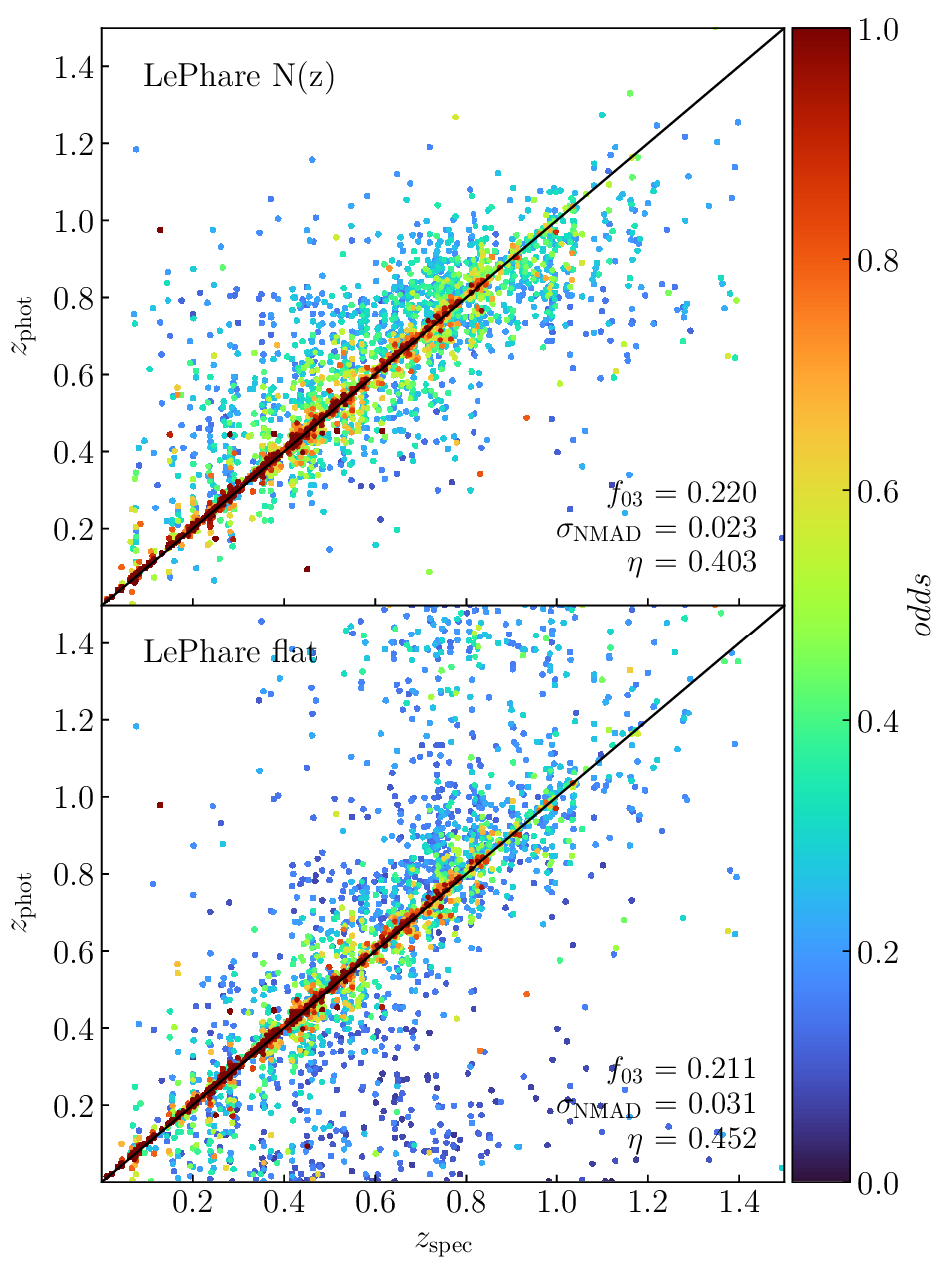}
\end{center}
\caption[]{Comparison of $z_{phot}$ versus $z_{spec}$ for individual galaxies in our sample. The value of $z_{phot}$ is obtained using the narrow bands of miniJPAS with the {\sc LePhare} code. The top panel shows results with a N($z$) prior applied to the likelihood, while the bottom panel shows results with a flat prior (see text for details). Symbols are colour-coded for the $odds$ parameter.\label{fig:zphot-zspec-JPAS}}
\end{figure}

\begin{figure} 
\begin{center}
\includegraphics[width=8.4cm]{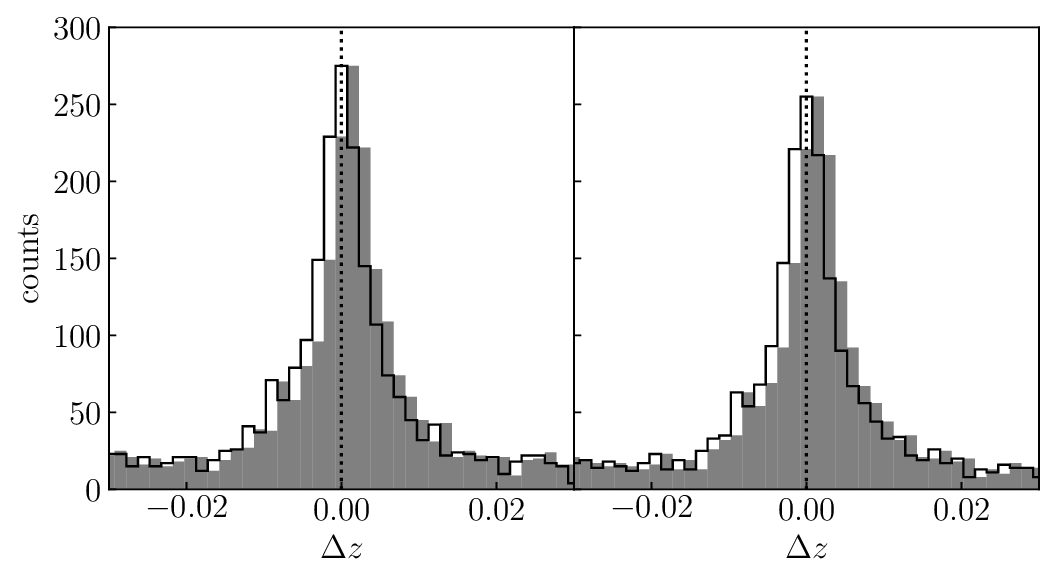}
\includegraphics[width=8.4cm]{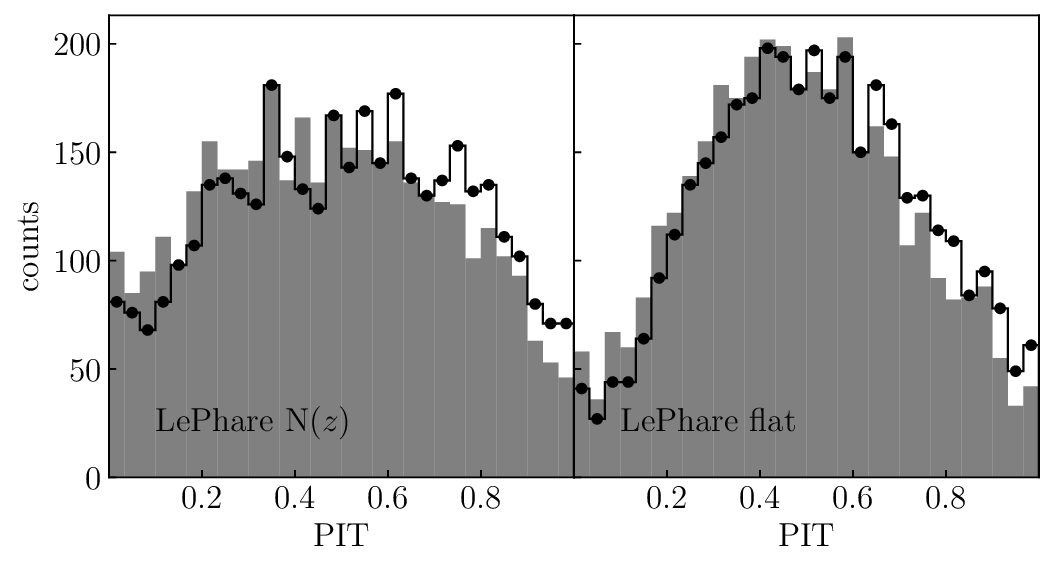}
\end{center}
\caption[]{Close-up of [-0.03,0.03] interval of the distribution of $\Delta z$ (top) and the distribution of PIT values (bottom) for photo-$z$ measurements with {\sc LePhare} using either a N($z$) prior (left) or a flat prior (right). Solid and open histograms represent the distributions obtained before and after correcting for a systematic offset $\delta z$=0.0015(1+$z$), respectively.\label{fig:dz-offset-LePhare}}
\end{figure}

The photo-$z$ in the Public Data Release of miniJPAS\footnote{\url{http://archive.cefca.es/catalogues/minijpas-pdr201912}} were obtained with the {\sc LePhare} code \citep{Arnouts11} using the 56 narrow bands of J-PAS in addition to the broader $u$, $g$, $r$, and $i$ bands. For this work, we recomputed the photo-$z$ using the same code and configuration but removing the $u$, $g,$ and $r$ bands to reproduce the actual survey strategy of J-PAS. 
We refer to \citet[][hereafter \citetalias{HC21}]{HC21} for details on the miniJPAS photometry and the photo-$z$ calculation. Very briefly, the photometry used here is obtained from the co-added miniJPAS images using {\sc SExtractor} in dual mode (the extraction aperture defined in the detection band is applied to all bands). We selected the PSF-corrected aperture (PSFCOR) 
photometry, which we corrected for Galactic extinction and for systematic offsets in the colour indices between J-PAS bands (see Sect. 3 in \citetalias{HC21}).
Then, we scaled the fluxes to match the flux in the larger AUTO aperture for the $i$ band. This maximizes the S/N and accuracy of the observed colour indices and prevents underestimation of the luminosity of the galaxies.
To estimate the photo-$z$, {\sc LePhare} computes the likelihood $\mathcal{L}$($z_i$) $\propto$ $\exp$[-$\chi^2_{min}$($z_i$)/2] for a discrete set of redshifts, $z_i$, where $\chi^2_{min}$ corresponds to the $\chi^2$ value of the best fitting template at redshift $z_i$. 
The templates are a set of 50 synthetic galaxy spectra generated with {\sc cigale} \citep{Boquien19}. The physical parameters that define the synthetic spectra (such as star formation history, metallicity, or extinction) are chosen by finding the values that best reproduce the observed photometry of individual galaxies in a small training sample. 

To compute the probability distribution for the redshift, $P$($z$), {\sc LePhare} modulates the raw likelihood with a redshift prior N($z$). The default prior for {\sc LePhare} is obtained from galaxy counts in the Vimos VLT Deep Survey \citep{LeFevre05}.   
The N($z$) prior helps break degeneracies in the colour-redshift space and improves the photo-$z$ accuracy for most galaxies, but it can also bias $z_{phot}$ values for populations at the faint end of the luminosity distribution \citep[such as dwarf elliptical galaxies; see Fig 18 in][]{Hernan-Caballero23}. To evaluate the impact of the N($z$) prior on the results, we also obtain photo-$z$ using a flat prior.

We take the mode of $P$($z$) as the best point estimate, $z_{phot}$. Figure \ref{fig:zphot-zspec-JPAS} shows $z_{phot}$ versus $z_{spec}$ for the two prior options. 
Using the N($z$) prior results in significantly lower $\sigma_{\rm{NMAD}}$ and $\eta$ compared to the flat prior because it removes solutions at the low and high extremes of the redshift search range that would correspond to unusually low or high luminosities, respectively. The impact of the prior on $f_{03}$ is very small, because obtaining $\vert\Delta z$$\vert$$<$0.003 typically implies a very narrow $\mathcal{L}$($z$), which is barely changed by the much broader N($z$) prior.

The top panels of Figure \ref{fig:dz-offset-LePhare} show the central part of the $\Delta z$ distribution. The main difference between the distributions with and without a prior is a slight increase in the number of sources with $\Delta z$$\sim$0 with the prior. An offset of $\delta z$ = 0.0015(1+$z$) is required in both cases to centre the distribution at $\Delta z$ = 0. 
The PIT diagram (bottom panel) shows a convex distribution, implying that the PDZs are too broad (under-confident). This under-confidence is stronger with the flat prior. Correcting for the $\delta z$ offset makes the PIT distributions more symmetric, but it does not decrease the curvature. 
There are 35 sources (less than 1\% of the sample) with PIT values of 0 or 1, which indicate $P$($z_{\rm{spec}}$) = 0.

\citet{HC21} also found under-confidence in the PDZs obtained with the 56 bands of J-PAS plus $u$, $g$, $r$, and $i$. They applied a power-law transformation to the PDZs to increase the contrast between peaks and valleys. This solved the under-confidence issue and resulted in roughly flat PIT distributions and realistic $odds$ estimates. 
However, since Figure \ref{fig:dz-offset-LePhare} shows that the choice of the prior impacts the PIT distribution and we intend to use the PDZs from HSC-SSP as a sort of new prior (different for each galaxy) with the conflation method, we refrained from applying contrast corrections to any of the PDZs.
We note that since the contrast correction does not change the mode of the PDZ, it does not affect in the value of $z_{phot}$. 

\section{Combined photometric redshifts}\label{sec:combination}

In this section, we describe and compare three methods for obtaining photo-$z$ with the combined HSC-SSP and J-PAS datasets: SED-fitting using the mixed photometry (broad bands and narrow bands fitted together with the same model), a weighted mean of point estimates from the two surveys, and the conflation of the two PDZs.

\begin{figure*}
\begin{center}
\includegraphics[width=18.0cm]{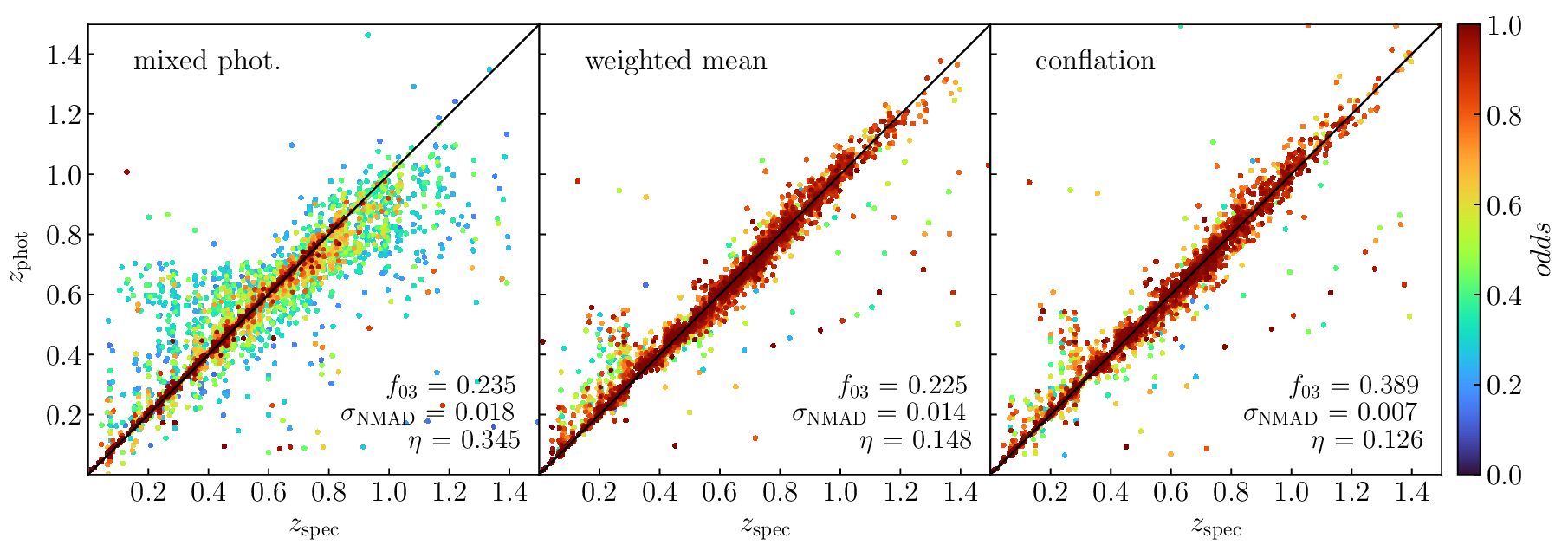}
\end{center}
\caption[]{Comparison of dispersion of $z_{phot}$ estimates obtained from the combined HSC-SSP + J-PAS dataset using SED-fitting of the mixed photometry (left panel), a weighted average of $z_{phot}$ values obtained independently in the two datasets (middle panel), and conflation of the probability distributions (right panel). The three panels contain the same sources. Symbols are colour-coded for the $odds$ parameter.\label{fig:zphot-zspec-methods}}
\end{figure*}

\subsection{SED-fitting the mixed photometry}

One of the main challenges of photo-$z$ estimation is ensuring that colour indices are accurate. While errors in the absolute flux calibration can introduce a systematic bias, the primary source of uncertainty is often the amount of flux lost outside the extraction aperture, which depends on the aperture size, the PSF of the image, and the light profile of the source. In point sources, aperture losses are easy to quantify and correct, but in extended sources, the uncertainty can be very considerable, especially if a small aperture is used to maximise the S/N of the photometry.

To decrease this uncertainty, photo-$z$ are often computed using aperture photometry extracted from images that are convolved to the same PSF in all the bands or using equivalent methods to compensate for PSF variation, such as PSFCOR. This approach can yield accurate colours if the differences in PSF FWHM are small and the light profile of the source does not change significantly from band to band.
More sophisticated methods that model both the light profile of the source and the PSF of the image, such as {\sc The Tractor} \citep{Lang16a,Lang16b}, {\sc The Farmer} \citep{Weaver23}, or {\sc SourceXtractor++} \citep{Bertin20,Kummel20} can overcome PSF variation without convolving the images.

Given that both miniJPAS and HSC-SSP already provided PSF-corrected photometry and that they have three bands in common ($g$, $r$, $i$), we chose a more straightforward strategy. We found a magnitude offset (different for each source) that when applied to the HSC-SSP bands minimises the average of the magnitude difference between miniJPAS and HSC-SSP in $g$, $r$, and $i$. 
From miniJPAS, we took the photometry described in Sect. \ref{sec:photoz-miniJPAS}.
From HSC-SSP, we took the photometry from the columns \textit{(g|r|i|z|y)\_convolvedflux\_2\_kron\_flux}
in the table \textit{pdr3\_wide.forced4}. This photometry is obtained using a Kron aperture \citep{Kron80} on images convolved to 1.1\arcsec{} FWHM (similarly to the typical FWHM of miniJPAS images) with the `afterburner' method \citep[see][for details]{Aihara18}. 
This ensures that the PSF FWHM and the aperture sizes are as close as possible to those in miniJPAS (the PSFCOR photometry of miniJPAS is also based on the Kron aperture; see Sect. 2.2 in \citetalias{HC21} for details).
\citet{Tanaka18} confirmed that photometry from convolved images provides the most accurate colours and photo-$z$ for both crowded and isolated objects in HSC-SSP. The most accurate photo-$z$ from HSC-SSP (obtained with the {\sc DEmP} code; see Sect. \ref{sec:photoz-subaru}) also use the Kron aperture on convolved images. 

We corrected the HSC-SSP photometry for Galactic extinction using the coefficients $a_x$ provided in the table \textit{pdr3\_wide.forced}. Then, for each source, we computed the magnitude offset that we applied to the HSC-SSP bands as the average colour difference: 
\begin{equation}
\delta m =  \frac{1}{3}[(g_H - g_J) + (r_H - r_J) + (i_H - i_J)] ,
\end{equation}
\noindent where the suffixes $H$ and $J$ correspond to HSC-SSP and miniJPAS, respectively.

After applying the offsets, we find a 1-$\sigma$ dispersion between HSC-SSP and miniJPAS magnitudes measured in the same band of 0.12, 0.08, and 0.09 mag for $g$, $r$, and $i$, respectively. This dispersion may be in part a consequence of differences in the transmission profiles of the filters.  

We computed photo-$z$ for the 61-band dataset (56 narrow-band filters of J-PAS plus $g$,$r$,$i$,$z$,$y$ from HSC-SSP) with {\sc LePhare} using the same configuration described in Sect. \ref{sec:photoz-miniJPAS} for J-PAS alone (including the N($z$) prior).
We added a quantity $\Delta m$ to the nominal errors of the HSC-SSP photometry to account for the systematic uncertainty between the J-PAS and HSC-SSP photometry. We tested the values $\Delta m$ = 0, 0.05, 0.1, and 0.2 mag. We obtain the best photo-$z$ accuracy for $\Delta m$ = 0.1 mag. The results are described in Sect. \ref{sec:compare-methods}. 

\subsection{Conflation of the PDZs}\label{sec:combined-zpdf}

The PDZs of J-PAS and HSC-SSP are represented by 32-bit floating point arrays containing the probabilities P($z_i$) sampled at a discrete set of redshifts \{$z_i$\}. Each survey uses a different \{$z_i$\}: from $z$ = 0 to $z$ = 1.5 in steps of 0.002 in J-PAS; from $z$ = 0 to $z$ = 6 in steps of 0.01 in HSC-SSP.

To compute the conflation of the two PDZs, we linearly interpolate the PDZ from HSC-SSP, $P_{H}$($z$), at the \{$z_i$\} of the J-PAS PDZ, $P_{J}$($z$). For this, we apply the systematic shifts $\delta z$=0.003(1+$z$) for HSC-SSP and $\delta z$=0.0015(1+$z$) for J-PAS found in Sect. \ref{sec:single-survey}.
Since, by construction $P_{J}$($z$$>$1.5) = 0, we discarded the $z$$>$1.5 range in $P_{H}$($z$). Then, Eq. \ref{eq:conflation} becomes
\begin{equation}\label{eq:combine-pdfs}
P_C(z) = \frac{P^-_{J}(z) \cdot P_{H}(z)}{\int_0^{zmax} P^-_{J}(z) P_{H}(z) dz} ,
\end{equation}
\noindent where $zmax$ = 1.5 and $P^-_J$($z$) is the version of the J-PAS PDZ obtained with a flat prior. The limited numerical precision implies that very small probabilities are rounded to zero. If the intervals where $P^-_J$($z$) $>$ 0 and $P_{H}$($z$) $>$ 0 do not overlap, then their product is always zero and $P_C$($z$) becomes undefined. In practice, zero overlap can happen only if one of the PDZs is unrealistically over-confident ($P$($z_{spec}$) = 0) or if the source association between the two surveys is wrong. None of the 3710 sources in our sample has undefined $P_C$($z$).

\subsection{Weighted mean of point estimates}

For an arbitrary PDZ, the expected value of the redshift, E($z$) = $\mu$, and its variance, $\sigma^2$, are given by 
\begin{equation}
\mu = \int_0^{zmax} z P(z) dz, \\
\sigma^2 = \Bigg{(} \int_0^{zmax} z^2 P(z) dz \Bigg{)} - \mu^2 .
\end{equation}
If the PDZ is Gaussian, it is entirely determined by the values of $\mu$ and $\sigma^2$.
Furthermore, the conflation of two Gaussian PDZs, $P_X$($z$) and $P_Y$($z$), is also Gaussian with the same mean and variance as the weighted mean of $\mu_X$ and $\mu_Y$ \citep{Hill11}:
\begin{equation}\label{eq:weighted-mean}
\mu_C = \frac{\sigma^2_Y \mu_X + \sigma^2_X \mu_Y}{\sigma^2_X + \sigma^2_Y}, \\
\sigma^2_C = \frac{\sigma^2_X \sigma^2_Y}{\sigma^2_X + \sigma^2_Y} .
\end{equation}
Therefore, assuming Gaussian PDZs, a weighted average of point estimates from miniJPAS and HSC-SSP would result in the same point estimates that we obtained from the conflated PDZs, with no need to use the PDZs.
Unfortunately, the PDZs are often far from Gaussian (especially at a low S/N) and the information on the position and relative strength of the multiple peaks is lost in the Gaussian approximation. As a consequence, using Eq. \ref{eq:weighted-mean} results in much lower photo-$z$ accuracy compared to PDZ conflation, as we show in the next section.

\subsection{Comparison of combination methods}\label{sec:compare-methods}

\begin{figure} 
\begin{center}
\includegraphics[width=8.4cm]{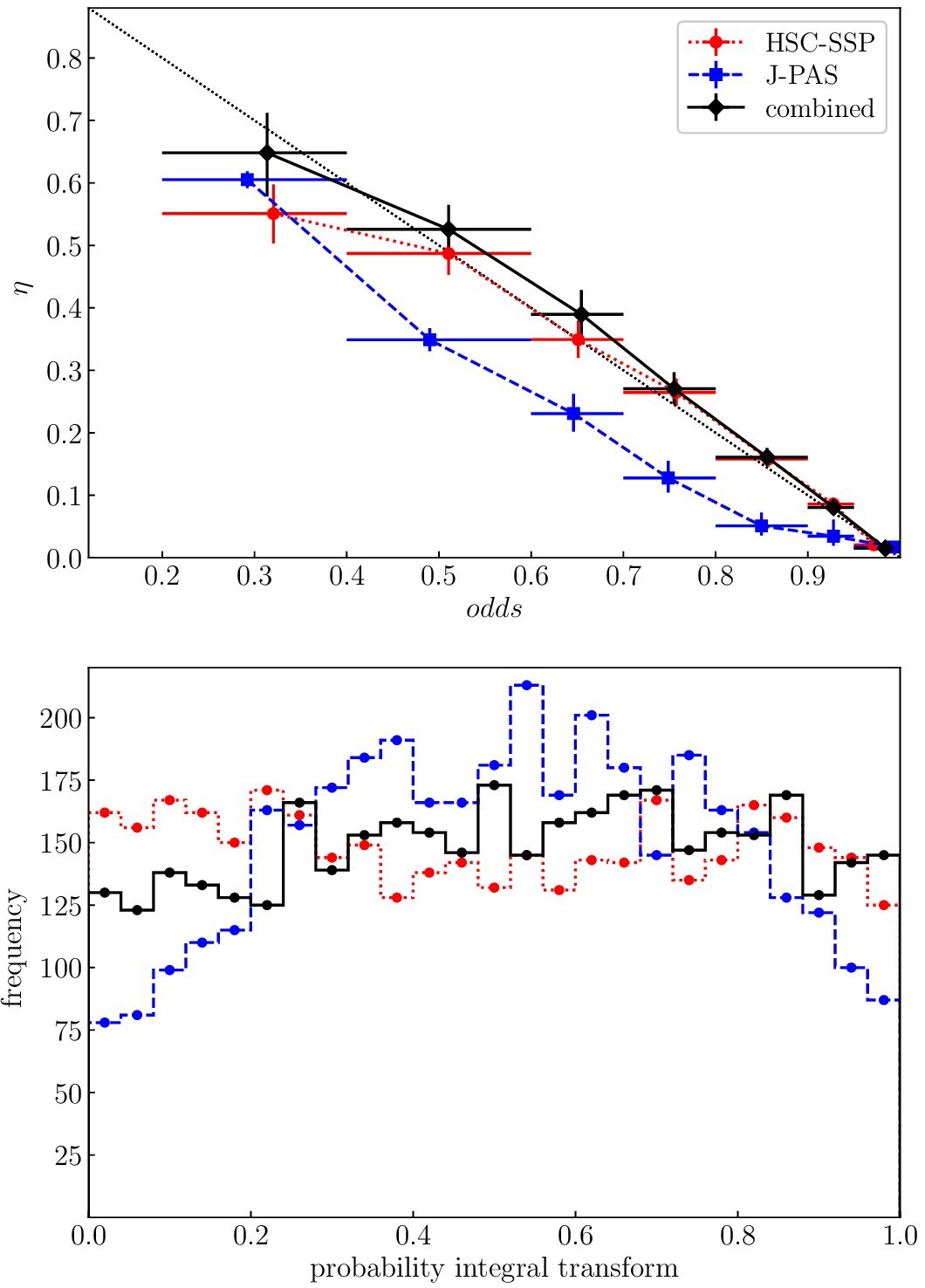}
\end{center}
\caption[]{Outlier rate ($\eta$) as a function of the $odds$ parameter (top panel) and the PIT test (bottom panel) using PDZs from HSC-SSP alone, J-PAS alone, and a combination of both with probability conflation. Error bars are as in Fig. \ref{fig:outrate-odds-SUBARU}.\label{fig:odds+PIT}}
\end{figure}

\begin{figure*}
\begin{center}
\includegraphics[width=18.0cm]{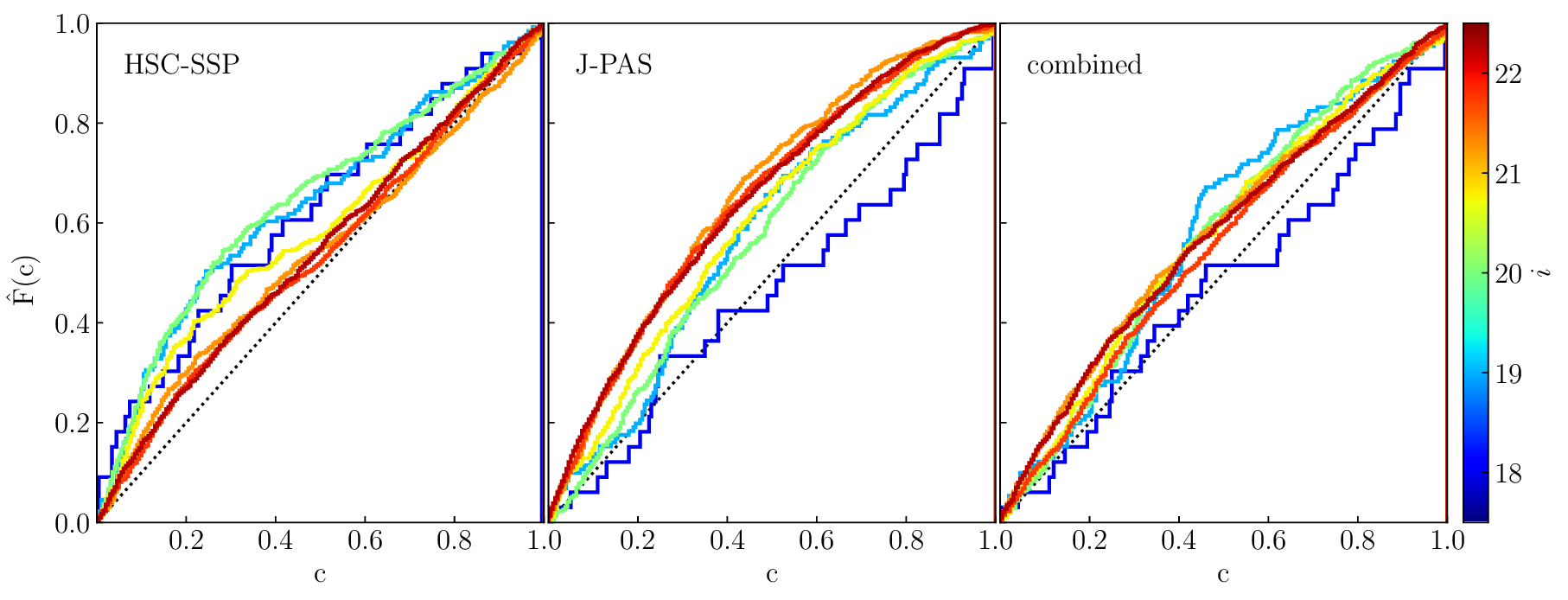}
\end{center}
\caption[]{Fraction of galaxies \^F($c$) for which $z_{spec}$ is inside the highest probability density confidence interval (HPDCI) as a function of the confidence level $c$, computed separately for the colour-coded magnitude bins $i$ = \{17.5--18.5, 18.5-19.5, 19.5--20.5, 20.5--21, 21--21.5, 21.5--22, 22--22.5\}. The left, middle, and right panels show results for the PDZs from HSC-SSP, J-PAS, and their conflation, respectively. Values above (below) the diagonal line indicate under-(over-)confidence in the PDZs.\label{fig:qqplot}}
\end{figure*}

Figure \ref{fig:zphot-zspec-methods} compares the dispersion of the $z_{phot}$ versus $z_{spec}$ relation obtained by the three methods for combining the HSC-SSP and J-PAS datasets. Table \ref{table:summary} summarises the accuracy statistics for all the datasets and methods. 

Out of the three combination methods, the mixed photometry is the worst performing, with scores only slightly better than those obtained from J-PAS alone and worse than HSC-SSP with {\sc DEmP} in $\sigma_{\rm{NMAD}}$ and $\eta$ (but not in $f_{03}$). 
This suggests that adding the HSC-SSP photometry does little to improve the photo-$z$ from the 56 bands of J-PAS, possibly due to the uncertainty in the magnitude offsets between the two datasets. It is likely that re-doing the photometry after convolving all the images to the same PSF would result in more accurate photo-$z$.

The weighted mean method results in roughly the same $f_{03}$ as the mixed photometry method or using J-PAS alone. However, both $\sigma_{\rm{NMAD}}$ and $\eta$ are much better. This suggests that the weighted mean is improving the photo-$z$ of faint sources, where the much larger uncertainty in the J-PAS $z_{phot}$ results in a mean value is closer to the HSC-SSP value (which is much less likely to be an outlier). For the bright sources, the uncertainty in the J-PAS $z_{phot}$ is small compared to HSC-SSP, therefore the solution is closer to the J-PAS value, which is usually the most accurate.

Finally, the PDZ conflation performs much better than the other two combination methods in $f_{03}$ and $\sigma_{\rm{NMAD}}$ (but only slightly better than the weighted mean method in $\eta$). PDZ conflation is also much better than either J-PAS or HSC-SSP separately in every score.
We find the fact that PDZ conflation is the only combination method that can significantly increase the fraction of galaxies with very small photo-$z$ errors with respect to what can be achieved with HSC-SSP or J-PAS alone particularly interesting. 
This establishes PDZ conflation as a simple yet powerful method for increasing the photo-$z$ accuracy for highly demanding tasks such as BAO measurements and highlights the importance of including the full PDZs in published photo-$z$ catalogues. 

\begin{figure*}
\begin{center}
\includegraphics[width=18.0cm]{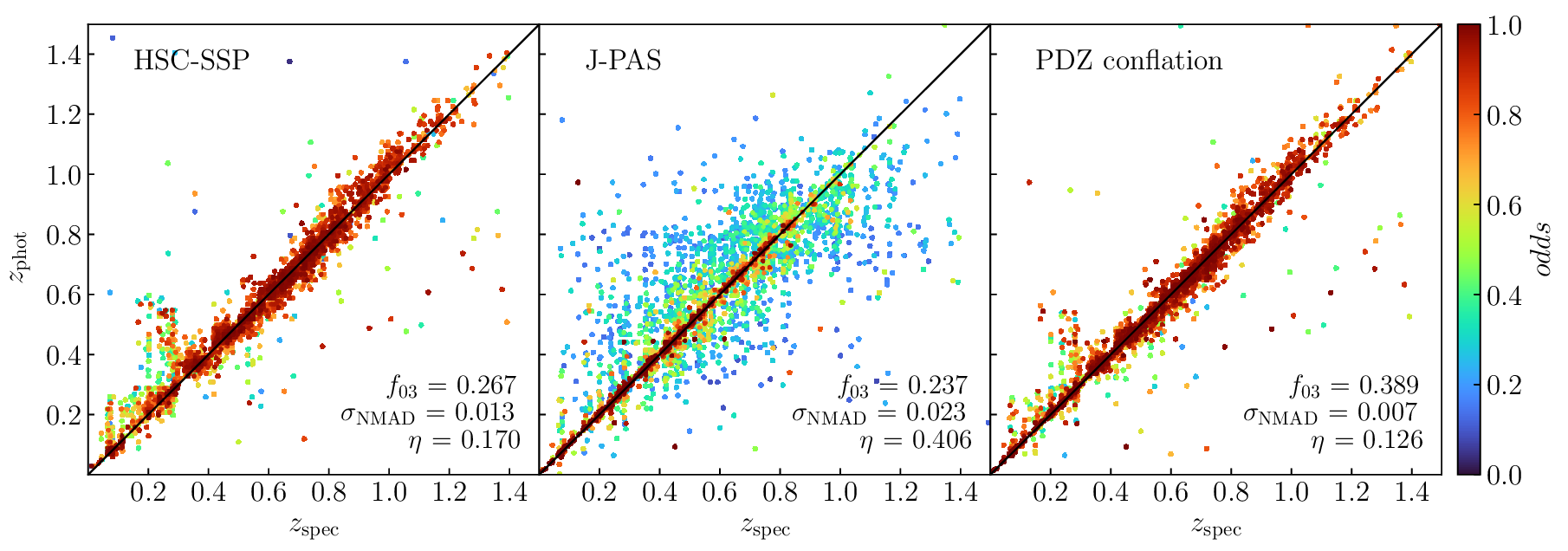}
\end{center}
\caption[]{Comparison between photometric and spectroscopic redshifts for individual galaxies in our sample. $z_{phot}$ marks the mode of the PDZ (corrected from systematic bias) from HSC-SSP (left), J-PAS narrow bands (middle), and the conflated PDZ of the two (right). Symbols are colour-coded for the $odds$ parameter.\label{fig:zphot-zspec-combined}}
\end{figure*}

\subsection{Validation of the conflated PDZs}

In the previous section, we established that point estimates with the conflation method are the most accurate. However, many applications also require confidence intervals derived from the PDZs be realistic. 
Figure \ref{fig:odds+PIT} shows the relation between $\eta$ and $odds$ and the PIT test for the PDZs from HSC-SSP with {\sc DEmP} (corrected for systematic offset), the narrow-band photometry of J-PAS ({\sc LePhare} code with flat prior, also corrected for systematic offset), and the conflation of the two PDZs.

The $odds$ values for J-PAS are strongly underconfident, while those for HSC-SSP correctly predict the outlier rate, except for the lowest $odds$ bin (0.2$<$$odds$$<$0.4), where they are also underconfident. On the other hand, the conflated PDZs are slightly overconfident for $odds$$>$0.4 (but consistent with the $\eta$ = 1 - $\langle odds \rangle$ relation within their 1-$\sigma$ uncertainties).
The PIT test (bottom panel in Fig. \ref{fig:odds+PIT}) shows a roughly flat distribution for the conflated PDZs, which suggests that the PDZs are, on average, realistic.
However, since the source counts increase steeply with the $i$-band magnitude, results from the PIT test are dominated by the faintest sources.

Another powerful test for the PDZs is the fraction of galaxies \^F($c$) in which $z_{spec}$ falls within a given confidence interval $c$ of the PDZ \citep[e.g.][]{Fernandez-Soto02,Dahlen13,Schmidt13}. For realistic PDZs, we expect \^F($c$) = $c$. Out of the many possible definitions of a confidence interval, the most useful is the highest probability density confidence interval (HPDCI), as proposed initially by \citet{Fernandez-Soto02} and illustrated by \citet{Wittman16}.
The HPDCI is the shortest interval (or union of disjoint intervals) that contains a given fraction of the total area under the PDZ distribution. Therefore, it always contains the mode of the PDZ.
\citetalias{HC21} obtained \^F($c$) versus $c$ relations for different magnitude bins, finding that, before applying a contrast correction, miniJPAS PDZs are slightly over-confident at bright magnitudes but increasingly under-confident at fainter ones. 

In Fig. \ref{fig:qqplot}, we show the same test on the PDZs from J-PAS, HSC-SSP, and their conflation. 
J-PAS PDZs present the same magnitude dependence found by \citetalias{HC21}, while HSC-SSP PDZs show the opposite trend: they are more under-confident at brighter magnitudes. Also, in the case of HSC-SSP, the arcs defined by the \^F($c$) versus $c$ relation are not symmetric with respect to the diagonal line, \^F($c$) = $c$, suggesting that the underconfidence is stronger for small values of $c$. Since the PDZs from HSC-SSP are usually unimodal, this implies that the probability density around the peak is underestimated at all magnitudes, but even more so for the brightest ones. This might be a consequence of the relatively coarse sampling (constant steps of 0.01 in $z$) of the PDZs published by the HSC Collaboration, which would be insufficient to resolve the actual shape of the PDZ near the peak.
Interestingly, the combination of the J-PAS and HSC-SSP PDZs results in \^F($c$) versus $c$ relations that are roughly symmetric with respect to \^F($c$) = $c$, with much less underconfidence compared to J-PAS alone, and with no clear magnitude dependence (except maybe for the brightest sources, where there is a hint of overconfidence that might be the result of small number statistics).
This is probably a consequence of the opposite magnitude dependencies in J-PAS and HSC-SSP cancelling each other out.

These tests indicate that conflating the PDZs from J-PAS and HSC-SSP results in realistic PDZs irrespective of the $odds$ and magnitude of the sources. Therefore, we consider that there is no need for an ad hoc contrast-correction of the J-PAS likelihood in this case. When conflating other datasets, a contrast correction of one or both PDZs might be required in order to prevent under- or over-confidence in the conflated PDZs. 

\section{Discussion and conclusions}\label{sec:discussion}

\subsection{Narrow-band versus deep broad-band photo-$z$}

\begin{figure} 
\begin{center}
\includegraphics[width=8.4cm]{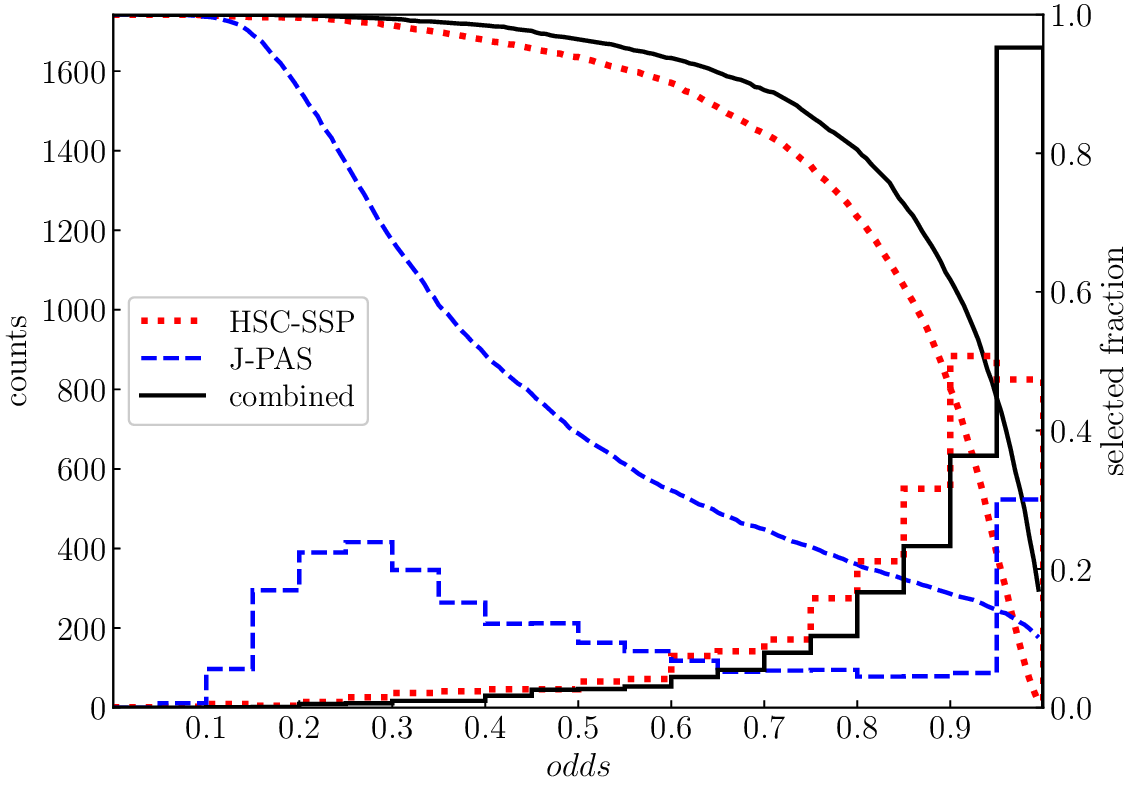}
\end{center}
\caption[]{Comparison of distribution of $odds$ values for $z_{phot}$ from HSC-SSP (red dotted lines), J-PAS (blue dashed lines), and the conflated PDZs (black solid lines). Histograms indicate galaxy counts in bins of $odds,$ while curved lines represent the fraction of the sample with $odds$ higher than a given threshold value (right axis).\label{fig:odds-distrib}}
\end{figure}

\begin{figure} 
\begin{center}
\includegraphics[width=8.4cm]{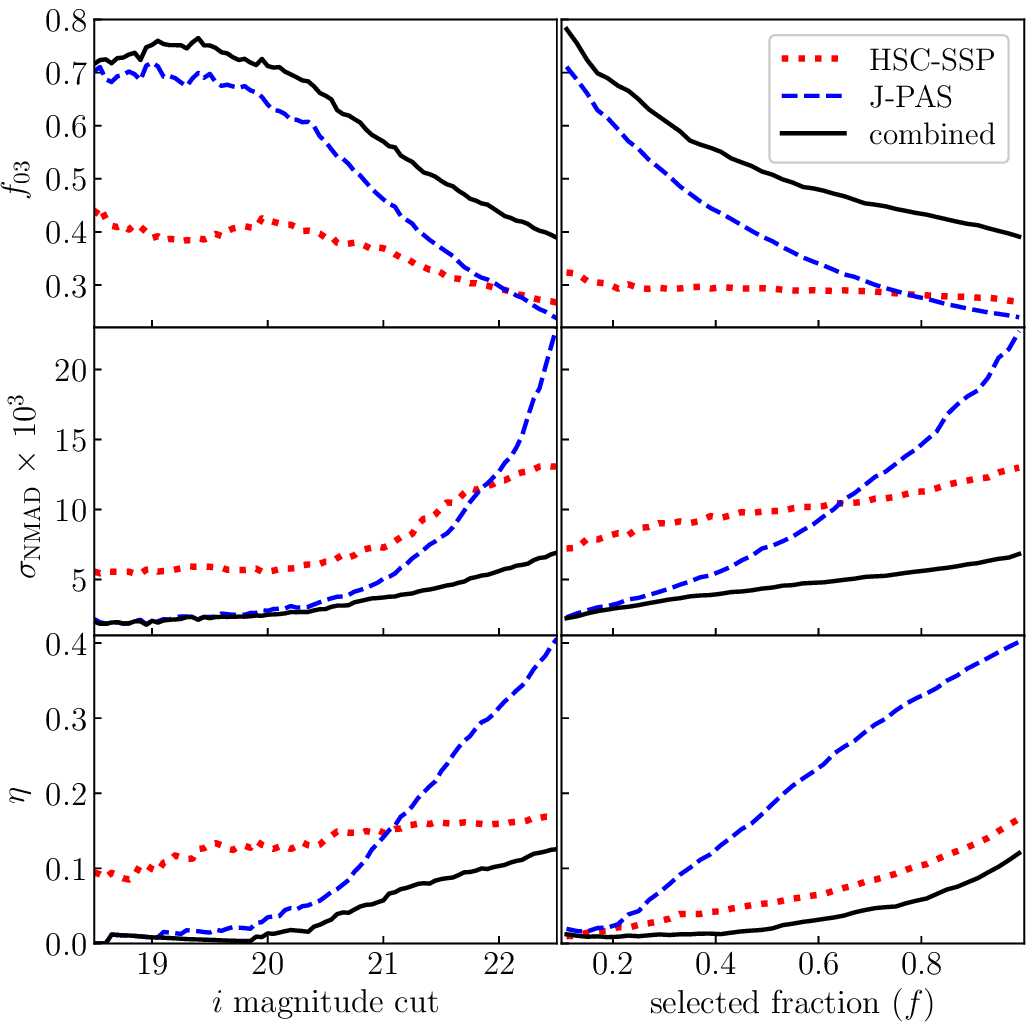}
\end{center}
\caption[]{Dependence of statistics $f_{\rm{03}}$, $\sigma_{\rm{NMAD}}$, and $\eta$ with the cut in magnitude applied (left panels) or with the fraction of the total sample selected using a cut in $odds$ (right panels).\label{fig:scores-selected-magcut}}
\end{figure}

The left and middle panels in Fig. \ref{fig:zphot-zspec-combined} show $z_{phot}$ versus $z_{spec}$ for $z_{phot}$ from HSC-SSP ({\sc DEmP} code) and J-PAS ({\sc LePhare} with N($z$) prior), respectively. Both were corrected for their respective systematic biases in $z_{phot}$. We note some remarkable differences between the two distributions. First, the dispersion for J-PAS is much higher, but it is almost entirely due to sources with very low $odds$ (blue-ish colours). On the other hand, high-$odds$ sources (red-ish colours) concentrate tightly around the 1:1 relation. In comparison, the dispersion of HSC-SSP photo-$z$ seems less dependent on the $odds$; albeit, it is hard to tell from this figure since most HSC-SSP photo-$z$ have high $odds$.

Our three main photo-$z$ accuracy statistics, $f_{03}$, $\sigma_{\rm{NMAD}}$, and $\eta$, are all better for HSC-SSP compared to J-PAS\footnote{We note that HSC-SSP outperforms J-PAS in $f_{03}$ only after correcting for the systematic bias in $z_{phot}$.}. This might seem surprising given the factor $\sim$ 8.5 increase in spectral resolution of J-PAS over HSC-SSP. However, this is entirely a consequence of pushing the magnitude limit of our sample selection down to $i$=22.5, which, given the red colours of most faint galaxies, implies that many sources have a very low S/N in most of the J-PAS narrow bands. The low S/N results in unreliable $z_{phot}$ estimates at faint magnitudes, as evidenced by the large fraction of sources with low J-PAS $odds$ (Fig. \ref{fig:odds-distrib}).

The statistics $f_{03}$, $\sigma_{\rm{NMAD}}$, and $\eta$ all show a much stronger dependence on the limiting magnitude for J-PAS compared to HSC-SSP (left panel in Fig. \ref{fig:scores-selected-magcut}). J-PAS obtains higher $f_{03}$ than HSC-SSP for $i$$<$22.1, lower $\sigma_{\rm{NMAD}}$ for $i$$<$21.9, and even lower $\eta$ for $i$$<$21. At bright magnitudes ($i$$\lesssim$20), both surveys saturate in their accuracy improvements, with J-PAS outperforming HSC-SSP by $\sim$80\%, $\sim$200\%, and $\sim$500\% in $f_{03}$, $\sigma_{\rm{NMAD}}$, and $\eta$, respectively. 
This indicates that, when considering magnitude-limited samples, narrow-band photometry can provide photo-$z$ with accuracy and robustness levels beyond the capabilities of broad-band photometry, as long as the S/N is not too low.
This also confirms that, for moderately bright galaxies ($i$$\lesssim$22), J-PAS on its own is competitive against much deeper (a factor $\sim$30 in S/N per band) broad-band surveys in obtaining complete samples with the highly accurate photo-$z$ needed for BAO studies.

\citet{HC21} proposed $odds$ cuts instead of magnitude cuts as a more efficient way to select sub-samples with accurate photo-$z$. We show the effect of an $odds$-based selection on the accuracy statistics in the right panels of Fig. \ref{fig:scores-selected-magcut}.
J-PAS overtakes HSC-SSP in $f_{03}$ when the fraction of the sample selected is $f$$<$0.8 and in $\sigma_{\rm{NMAD}}$ for $f$$<$0.63. However, even very strong $odds$ cuts do not achieve a lower $\eta$ than HSC-SSP at the same $f$ (both converge towards $\eta$ = 0 at $odds$ = 1, as expected). 

These trends reflect the different nature of the uncertainties in the broad-band and narrow-band photo-$z$, which are a consequence of the shape of their PDZs (see Fig. \ref{fig:pdf-examples}); broad-band photo-$z$ provide highly confident (low $\eta$) estimates thanks to the high S/N of the photometry, but with little accuracy due to the limited spectral resolution (therefore, the PDZs are broad and unimodal). On the other hand, narrow-band photo-$z$ provide higher accuracy due to higher spectral resolution but with lower confidence given the low S/N (the PDZs contain multiple narrow peaks).

\subsection{Photo-$z$ accuracy from the conflated PDZs}\label{sec:accuracy-combined}

\begin{figure} 
\begin{center}
\includegraphics[width=8.4cm]{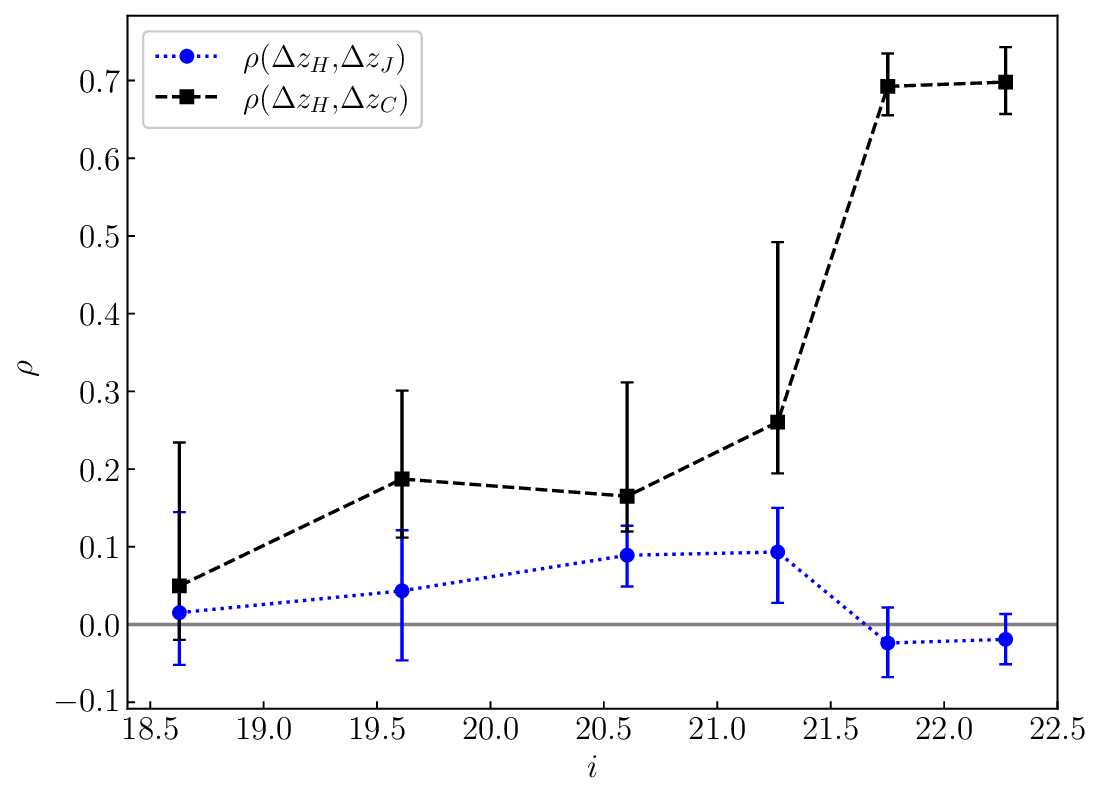}
\end{center}
\caption[]{Pearson correlation coefficient between $\Delta z$ values for $z_{phot}$ measured on HSC-SSP and conflated PDZs (black squares) and between HSC-SSP and J-PAS (blue circles) as a function of the $i$-band magnitude. Error bars represent 1-$\sigma$ confidence intervals obtained with bootstrap resampling.\label{fig:pearson-test}}
\end{figure}

We show $z_{phot}$ computed as the mode of the conflated PDZ versus $z_{spec}$ in the right panel of Fig. \ref{fig:zphot-zspec-combined}. Visually, the distribution looks similar to that of HSC-SSP, albeit with less dispersion, particularly at low $z$. 
The improvement in the scores is substantial: $\sim$50\% more sources with $\vert\Delta z\vert$$<$0.3\%, a factor $\sim$2 decrease in $\sigma_{\rm{NMAD}}$ compared to HSC-SSP (factor $\sim$3 compared to J-PAS), and an $\sim$25\% decrease in the outlier rate compared to HSC-SSP ($\sim$70\% compared to J-PAS). 

\begin{table}[] 
\setlength{\tabcolsep}{3.5pt}
\small
\caption{Summary of accuracy statistics for point estimates.\label{table:summary}} 
\begin{center}
\begin{tabular}{c|cccccc} 
\hline
\hline
Dataset & Method & $f_{03}$ & $f_1$ & $\sigma_{\rm{NMAD}}$ & $\eta$ & $\eta_{15}$ \\
\hline
\multirow{4}{*}{HSC-SSP} & {\sc Mizuki} & 0.096 & 0.303 & 0.025 & 0.294 & 0.039 \\
 & {\sc DNNz} & 0.111 & 0.346 & 0.022 & 0.247 & 0.036 \\
 & {\sc DEmP} & 0.196 & 0.539 & 0.013 & 0.173 & 0.025 \\
 & {\sc DEmP}$^{c}$ & 0.267 & 0.533 & 0.013 & 0.170 & 0.024 \\
\hline
\multirow{3}{*}{J-PAS} & {\sc LePhare} flat & 0.211 & 0.409 & 0.031 & 0.452 & 0.160 \\
 &  {\sc LePhare} N($z$) & 0.220 & 0.436 & 0.023 & 0.403 & 0.079 \\
 & {\sc LePhare} N($z$)$^{c}$ & 0.237 & 0.437 & 0.023 & 0.406 & 0.079 \\
\hline
\multirow{4}{*}{combined} &          mixed phot. & 0.235 & 0.467 & 0.018 & 0.345 & 0.057 \\
 &   w. mean flat & 0.225 & 0.522 & 0.014 & 0.149 & 0.015 \\
 & w. mean N($z$) & 0.225 & 0.527 & 0.014 & 0.148 & 0.018 \\
 &  \textbf{conflation} & \textbf{0.389} & \textbf{0.671} & \textbf{0.007} & \textbf{0.126} & \textbf{0.015} \\
\hline
\end{tabular}
\end{center}
$^c$ corrected from systematic offset in $z_{phot}$.
\end{table}

Figure \ref{fig:scores-selected-magcut} shows that the conflation of the HSC-SSP and J-PAS PDZs improves all the scores, not only for the whole $i$$<$22.5 sample, but also for any other magnitude cut or selected fraction using an $odds$ cut. At bright magnitudes, the conflation results converge to those of J-PAS alone. 
For fainter sources, results are expected to eventually converge towards the HSC-SSP values as the S/N in the J-PAS bands approaches zero.
This is confirmed by Figure \ref{fig:pearson-test}, which shows a steep increase with magnitude in the correlation between errors in $z_{phot}$ from HSC-SSP and the conflated PDZs. However, even in the faintest magnitude bin (22.0$<$$i$$<$22.5), we obtain $\rho$$<$1, which is indicative of some contribution from the J-PAS PDZ to the $z_{phot}$ with conflation.
On the other hand, the correlation between errors in $z_{phot}$ values for HSC-SSP and J-PAS is consistent with $\rho$=0 at all magnitudes, confirming that the assumption of independence between PDZs that is central to the conflation method is valid in this case.

We interpret the big improvement in photo-$z$ accuracy with conflation as evidence for the complementarity between shallow narrow-band and deep broad-band surveys for photo-$z$.  
By conflating the J-PAS and HSC-SSP PDZs, we increase the intensity of those peaks in the J-PAS PDZ that are consistent with the HSC-SSP PDZ while filtering the spurious peaks that are inconsistent. We anticipate that improvements from PDZ conflation will be less dramatic when combining datasets with the same depth or spectral resolution. 

The improvement in photo-$z$ accuracy from the conflated PDZs over using J-PAS alone can also be expressed in terms of the survey depth at the same photo-$z$ accuracy.
The statistics obtained with conflation for a magnitude cut $i$$<$22.5 are only achieved by J-PAS alone at $i$$<$21 in the case of $\eta$ or $i$$<$21.3 for $\sigma_{\rm{NMAD}}$ and $f_{03}$. In other words, the conflation of the PDZs provides photo-$z$ accuracy comparable to increasing the depth of J-PAS by $\sim$1.2--1.5 magnitudes.

\begin{acknowledgements}

We thank the anonymous referee for their valuable comments and suggestions. 
This paper is based on observations made with the JST/T250 telescope at the Observatorio Astrof\'isico de Javalambre (OAJ) in Teruel, owned, managed, and operated by the Centro de Estudios de F\'isica del Cosmos de Arag\'on (CEFCA). We acknowledge the OAJ Data Processing
and Archiving Unit (UPAD) for reducing and calibrating the OAJ data used in this work.
Funding for the J-PAS Project has been provided by the Governments of Spain and Arag\'on through the Fondo de Inversi\'on de Teruel, European FEDER funding and the Spanish Ministry of Science, Innovation and Universities, and by the Brazilian agencies FINEP, FAPESP, FAPERJ and by the National Observatory of Brazil. Additional funding was also provided by the Tartu Observatory and by the J-PAS Chinese Astronomical Consortium.
Funding for OAJ, UPAD, and CEFCA has been provided by the Governments of Spain and Arag\'on through the Fondo de Inversiones de Teruel; the Arag\'on Government through the Research Groups E96, E103, and E16\_17R; the Spanish Ministry of Science, Innovation and Universities (MCIU/AEI/FEDER, UE) with grant PGC2018-097585-B-C21; the Spanish Ministry of Economy and Competitiveness (MINECO/FEDER, UE) under AYA2015-66211-C2-1-P, AYA2015-66211-C2-2, AYA2012-30789, and ICTS-2009-14; and European FEDER funding (FCDD10-4E-867, FCDD13-4E-2685). A.H.-C. and J.A.F.O. acknowledge financial support by the Spanish Ministry of Science and Innovation (MCIN/AEI/10.13039/501100011033) and ``ERDF A way of making Europe'' through the grant PID2021-124918NB-C44. J.A.F.O. acknowledges funding by MCIN and the European Union -- NextGenerationEU through the Recovery and Resilience Facility project ICTS-MRR-2021-03-CEFCA. C.H.-M. acknowledges the support of the Spanish Ministry of Science and Innovation through project PID2021-126616NB-I00. J.C.M. acknowledges support from the European Union’s Horizon Europe research and innovation programme (COSMO-LYA, 101044612). L.A.D.G. acknowledges financial support from the State Agency for Research of the Spanish MCIU through `Center of Excellence Severo Ochoa' award to the Instituto de Astrof\'isica de Andaluc\'ia (SEV-2017-0709) and CEX2021-001131-S funded by MCIN/AEI/10.13039/501100011033 and from the project PID-2019-109067-GB100.
A.F.-S. acknowledges support by project PID2019-109592GBI00/AEI/10.13039/501100011033 from the Spanish Ministerio de Ciencia e Innovaci\'on (MCIN)—Agencia Estatal de Investigaci\'on, by the Project of Excellence Prometeo/2020/085 from the Conselleria d’Innovaci\'o Universitats, Ci\`encia i Societat Digital de la Generalitat Valenciana, and by the MCIN with funding from the European Union-NextGenerationEU and Generalitat Valenciana in the Programa de Planes Complementarios de I+D+i (PRTR 2022) Project (VAL-JPAS), reference ASFAE/2022/025.
A.dP. acknowledges the financial support from the European Union - NextGenerationEU and the Spanish Ministry of Science and Innovation through the Recovery and Resilience Facility project J-CAVA and the project PID2021-124918NB-C41.
A.L.-C. acknowledges funding by the European Union - NextGenerationEU through the Recovery and Resilience Facility program Planes Complementarios con las CCAA de Astrof\'{\i}sica y F\'{\i}sica de Altas Energ\'{\i}as - LA4. R.G.D. acknowledges financial support from the grants CEX2021-001131-S and PID-2019-109067-GB100 funded by MCIN/AEI/10.13039/501100011033. J.M.V. acknowledges financial support from the grant PID2019-107408GB-C44 funded by MCIN/AEI/10.13039/501100011033.
P.C. acknowledges support from Conselho Nacional de Desenvolvimento Cient\'ifico e Tecnol\'ogico (CNPq) under grant 310555/2021-3 and from Funda\c{c}~{a}o de Amparo `{a} Pesquisa do Estado de S~{a}o Paulo (FAPESP) process number 2021/08813-7. Y.J-T. acknowledges financial support from the European Union’s Horizon 2020 research and innovation programme under the Marie Sk\l{}odowska-Curie grant agreement No 898633, the MSCA IF Extensions Program of the Spanish National Research Council (CSIC), the State Agency for Research of the Spanish MCIU through the Center of Excellence Severo Ochoa award to the Instituto de Astrofísica de Andalucía (SEV-2017-0709), and grant CEX2021-001131-S funded by MCIN/AEI/ 10.13039/501100011033
P.A.A.L. thanks the support of CNPq (grants 433938/2018-8 e 312460/2021-0) and FAPERJ (grant E-26/200.545/2023).
E.T. acknowledges the support by ETAg grant PRG1006.

\end{acknowledgements}

\appendix

\section{Conflation with {\sc Mizuki} and {\sc DNNz} PDZs}\label{sec:appendix}

\begin{figure*}[!t]
\begin{center}
\includegraphics[width=18.0cm]{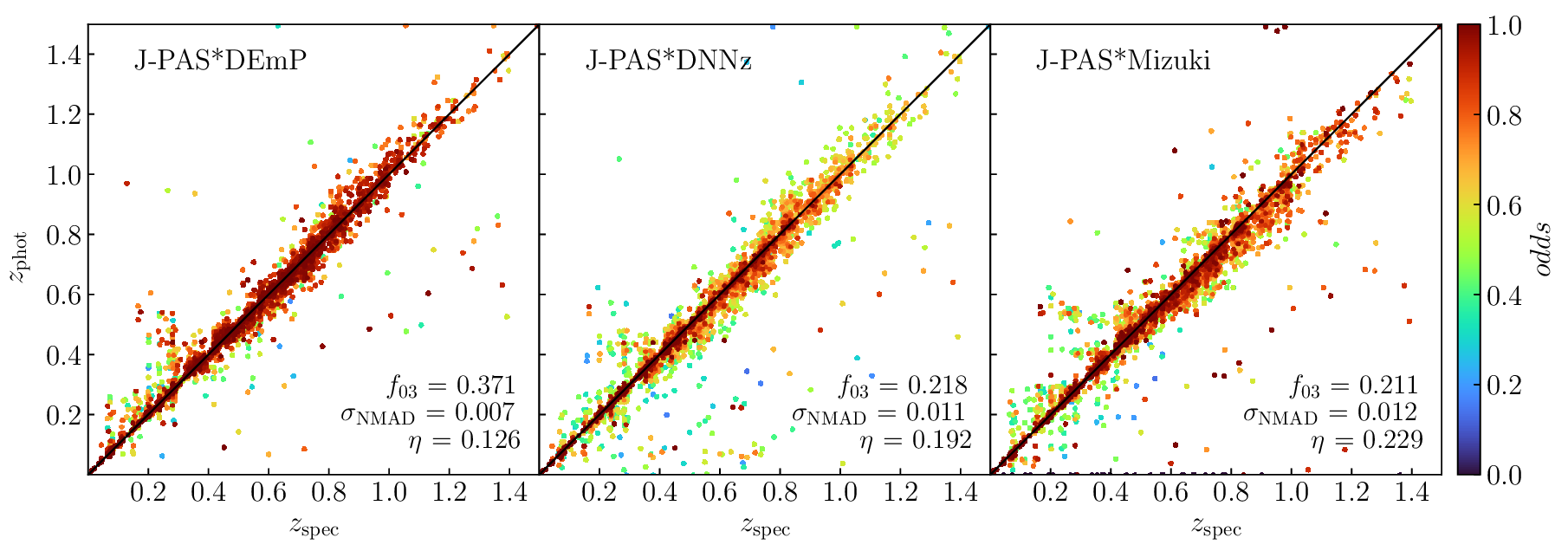}
\end{center}
\caption[]{Comparison between $z_{\rm{phot}}$ and $z_{\rm{spec}}$ for the galaxies in our sample, with $z_{\rm{phot}}$ obtained from the conflation of the J-PAS likelihood with the PDZs from HSC-SSP obtained with the {\sc DEmP} (left), {\sc DNNz} (middle), and {\sc Mizuki} (right) codes. Symbols are colour-coded for the $odds$ parameter.\label{fig:zphot-zspec-conflated3}}
\end{figure*}

In this paper, we focus on the results of combining the J-PAS likelihood with the PDZ from {\sc DEmP} since this is the code that provides the most accurate photo-$z$ from HSC-SSP observations (see Sect. \ref{sec:photoz-subaru}). However, our conclusions on the benefits of conflation are not specific to a particular photo-$z$ code or dataset. In this appendix, we compare the results of conflation between the J-PAS likelihood and the PDZs from {\sc DEmP}, {\sc Mizuki,} and {\sc DNNz}.

To obtain the conflated PDZs, we follow the procedure described in Sect. \ref{sec:combined-zpdf}, but in this case, we do not try to correct the systematic redshift offsets in the PDZs from HSC-SSP (we assume $\delta z$ = 0). This results in a slightly lower $f_{03}$ for the conflation with {\sc DEmP} PDZs compared to previous results with a $\delta z$ correction ($f_{03}$ = 0.371 instead of 0.389), but the impact on $\sigma_{\rm{NMAD}}$ and $\eta$ is negligible.

Figure \ref{fig:zphot-zspec-conflated3} compares the accuracy of $z_{phot}$ from the conflated PDZs for the three codes. The results obtained with {\sc DEmP} are significantly more accurate than those from the other two codes, while {\sc DNNz} is only marginally more accurate than {\sc Mizuki}. This replicates the relative performance of the codes on the HSC-SSP photometry alone (Sect. \ref{sec:photoz-subaru}), confirming our expectation that improving the photo-$z$ accuracy in one dataset translates into improved accuracy from the conflation. 
We note that regardless of the photo-$z$ code chosen, the results from conflation are significantly better compared to J-PAS or HSC-SSP alone.

\end{document}